\documentclass[nofootinbib,superscriptaddress,12pt]{revtex4}

\usepackage[usenames]{color}
\usepackage[colorlinks=true
,urlcolor=blue
,anchorcolor=blue
,citecolor=green
,filecolor=blue
,linkcolor=blue
,menucolor=blue
,pagecolor=blue
,linktocpage=true
]{hyperref}

\usepackage[a4paper,lmargin=70pt,rmargin=70pt]{geometry}
\usepackage{amsmath}
\usepackage{amssymb}
\usepackage{graphicx}
\usepackage{subfigure}
\usepackage{float}

\newcommand{\ed}{\,.}
\newcommand{\ec}{\,,}
\newcommand{\ecq}{\ec\quad}

\newcommand{\Pf}{\text{Pf}}
\newcommand{\as}{\alpha_s}
\newcommand{\GeV}{\mathrm{GeV}}
\newcommand{\TeV}{\mathrm{TeV}}

\newcommand{\ttf}{2 \to 4}
\newcommand{\ttt}{2 \to 2}

\DeclareMathOperator{\trace}{Tr}

\newcommand{\cC}{\ensuremath{\mathcal{C}}}
\newcommand{\cM}{\ensuremath{\mathcal{M}}}
\newcommand{\cO}{\ensuremath{\mathcal{O}}}

\newcommand{\cT}{\ensuremath{\mathcal{T}}}
\newcommand{\cJ}{\ensuremath{\mathcal{J}}}
\newcommand{\cI}{\ensuremath{\mathcal{I}}}

\def\sect#1{section~\ref{#1}}

\def\pT{p_{\rm T}}
\def\kT{k_{\rm T}}
\def\tbar{\bar t}
\def\Sherpa{SHERPA}

\begin{document}

\begin{flushright}
\vspace*{-.5cm}\tiny CERN-PH-TH/2012-318
\end{flushright}

\title{Three-Prong Distribution of Massive Narrow QCD Jets}

\author{Matan Field} \affiliation{Harish-Chandra Research Institute, \\
Chhatnag Road, Jhusi, Allahabad 211019, India
}

\author{Guy Gur-Ari} \affiliation{Department of Particle Physics \&
Astrophysics, Weizmann Institute of Science, Rehovot 76100, Israel}

\author{David A. Kosower} \affiliation{Institut de Physique Th\'eorique, CEA-Saclay, F-91191 Gif-sur-Yvette cedex, France}

\author{Lorenzo Mannelli} \affiliation{Department of Particle Physics \&
Astrophysics, Weizmann Institute of Science, Rehovot 76100, Israel}

\author{Gilad Perez} \affiliation{Department of Particle Physics \&
Astrophysics, Weizmann Institute of Science, Rehovot 76100, Israel}
\affiliation{CERN Physics Department, Theory Division, CH-1211 Geneva 23, Switzerland}

\vspace*{.5cm}

\setlength{\parindent}{0pt}
\setlength{\parskip}{1ex}

\begin{abstract}
\vspace*{.15cm}
  We study the planar-flow distributions of narrow,
  highly boosted, massive QCD jets. Using the factorization properties
  of QCD in the collinear limit, we compute the planar-flow jet
  function from the one-to-three splitting function at tree-level. We
  derive the leading-log behavior of the jet function analytically.
  We also compare our semi-analytic jet function with parton-shower
  predictions using various generators.
\end{abstract}

\maketitle
\newpage
\tableofcontents

\section{Introduction}

Observables sensitive to the substructure of energetic, ultra-massive jets
hold great promise for distinguishing new physics signals from QCD
backgrounds.  Both the ATLAS and CMS experiments
are pursuing studies relying on such observables 
in various new physics searches~\cite{:2012txa,Chatrchyan:2012cx} (see also Refs.~\cite{Aad:2012meb,Rappoccio:2012zz}, and Ref.~\cite{Aaltonen:2011pg} for an earlier search by the CDF
collaboration). Boosted jets originating from electroweak gauge
bosons~\cite{Butterworth:2002tt}, top
quarks~\cite{Agashe:2006hk,Lillie:2007yh}, Higgs
bosons~\cite{Butterworth:2008iy}, and even new physics
particles~\cite{Butterworth:2009qa,Butterworth:2007ke} are all of
interest as targets of searches by the Tevatron and the LHC
experiments.  It is therefore important to be able to distinguish them
from QCD jets.  Recent reviews on substructure techniques,
experimental status and new physics searches include
Refs.~\cite{Ellis:2007ib,Abdesselam:2010pt,Salam:2009jx,Nath:2010zj,Almeida:2011ud}
and references therein.

One way to characterize jet substructure is to consider observables which are functions of the energy flow within the jet, namely the energy distribution as measured by the detector (see Ref.~\cite{GurAri:2011vx} for a recent systematic classification). In this work we consider the hadronic collision
\begin{align}
  H_A + H_B \to J(\cO;\pT,\eta;R) + X \ec
\end{align}
where $H_A,H_B$ are the initial hadrons, and $J$ is a jet with momentum given by $\pT$ and $\eta$, with size $R$ determined by the jet algorithm,
 and characterized by an energy-flow observable $\cO$ such as the jet mass $m$. Note that $\cO$ can stand for multiple energy-flow observables. We focus on narrow, highly-boosted jets, and consider the inclusive differential QCD cross section for this process,
\begin{align}
  \frac{d\sigma_{2 \to JX}(R)}{d\pT \, d\eta \, d\cO} \ed
  \label{cs}
\end{align}
Such cross sections are usually computed using parton-shower codes, which offer much less insight into results than analytic computations.
They also require significant computational resources.

In this article we follow a different approach, based on the collinear
factorization properties of amplitudes in perturbative, massless
QCD. Factorization allows us to focus on a single jet, ignoring
to leading order the rest of process, and to compute the differential
jet substructure distribution semi-analytically.  In some
limiting cases, we can compute the distribution
completely analytically. As we demonstrate below, this method
becomes useful when analyzing jets with sufficiently high $\pT$
($\pT\gtrsim 1$ TeV), a window recently opened at the LHC (see for example
Refs.~\cite{:2012txa, Chatrchyan:2012cx}).

\def\Ord{{\cal O}} 

There are various ways to define jet shapes. In the context of
new-physics searches, a particularly useful way to characterize jet
shape and substructure observables is by the first non-trivial order
at which they appear in fixed-order perturbation
theory~\cite{Almeida:2008yp,GurAri:2011vx,Almeida:2010pa}.  Consider, for instance, Higgs
boson searches: at leading order the Higgs boson decays
to two partons with large invariant mass. The relevant non-trivial
substructure observables must distinguish two-prong jets from the
broad spectrum of all jets. Given a Higgs-boson mass and $\pT$, its
decay kinematics are fully determined by one additional continuous
variable, such as the ratio between the two decay-product
momenta~\cite{Butterworth:2008iy} or the opening angle between
them~\cite{Almeida:2008yp}.  The first non-trivial QCD background
arises from corrections in which the jet is made up of two partons.
If the Higgs boson recoils against a jet, this background first arises
at next-to-leading order (NLO) in two-jet production, when
$2\rightarrow3$ processes are taken into account.  This contribution
is of $\Ord(\alpha_s^3)$ in fixed-order perturbation theory.

In this example, the signal distribution at fixed $\pT$ is fully
characterized by the jet mass and jet angularity~\cite{Berger:2003iw,Almeida:2008yp}.  We may think of
these two quantities as replacing partonic (and therefore
unphysical) parameters with physical infrared- and collinear- (IRC) safe jet-shape
observables~\cite{Almeida:2008yp}.  When working to leading order (LO)
in the jet mass, the angularity distribution is the only independent
jet-shape observable that can separate the signal from the background.
The corresponding LO distribution, given a mass cut, can be computed
analytically both for the signal and background~\cite{Almeida:2008yp},
using the collinear approximation, which is adequate for narrow massive
jets.  The difference between the signal and background distributions
turns out to be modest, because both the QCD and Higgs-boson angularity
distributions are monotonically decreasing functions between identical
limiting values of the angularity variable.  The same result obtains
when considering the ratio of momenta or other kinematical variables,
because they are all fully correlated with the angularity distribution.  The similarity of
the bounds on both the signal and background does yield a sharp
prediction of this apparently naive picture.  This prediction was
qualitatively verified experimentally in the CDF collaboration's study~\cite{Aaltonen:2011pg}
of high-$\pT$ massive jets.

The next example of interest involves studies of high-$\pT$ top
quarks.  At leading order, each top quark decays to three partons; in
the decay, each parton pair typically
has a large invariant mass.  The same configuration is also relevant
to studies of new physics~\cite{Agashe:2006hk, Lillie:2007yh}, for
example of gluino decay in $R$-parity violating scenarios~\cite{Brooijmans:2010tn}.  A useful jet-shape observable in such a study is
the planar flow~\cite{Almeida:2008yp,Almeida:2008tp,Thaler:2008ju}.  If we focus on
studying one top quark out of the produced pair, the leading QCD
background is two-jet production where we constrain one of the jets to
have a significant planar flow.  This background first arises at
next-to-next-to-leading order (NNLO) in two-jet production, when
$2\rightarrow4$ processes are taken into account.  The contribution is
of $\Ord(\alpha_s^4)$ in fixed-order perturbation theory.

While the planar-flow distribution of a top-quark jet at leading order can be computed straightforwardly from its known matrix element,
the planar-flow distribution 
of the QCD background has not yet been computed for narrow massive jets. The corresponding distribution was presented in Ref.~\cite{Aaltonen:2011pg} with a rather limited sample size due to the limited statistics of massive boosted jets at the Tevatron.  LHC experiments have collected a much larger number of massive jets,
which should allow a more precise measurement of the planar-flow distribution.

Our main purpose in this article is to compute the planar-flow
distribution of narrow massive jets to leading order in QCD.  We use
the collinear approximation, in which we approximate the matrix
element for such jets by $1\rightarrow3$ collinear splitting
functions.  Motivated by boosted-top studies, we take the jet mass to be roughly
the top-quark mass.  Our approximation is relevant when the jet $\pT$ is
substantially larger than the jet mass; we take the $\pT$ to be
$\Ord(1~\TeV)$. Our computation assumes a jet algorithm that produces approximately circular jets with radius $R$
in the pseudorapidity--azimuthal~angle plane, but is otherwise general.  
The use of the collinear approximation also requires that the jet 
radius not be too large; we take $R=0.4$. In parton-shower simulations to which we compare,
we use the anti-$\kT$ algorithm, with the same jet size. As already mentioned, this year's $t\tbar$-resonance searches are already
exploring this range of parameter space.

Jet shape observables can be viewed as moments of the energy distribution within a jet~\cite{GurAri:2011vx}.
They are highly susceptible to contamination from pile-up and other sources of soft radiation, 
especially for larger cone sizes~\cite{Cacciari:2008gn, Krohn:2009wm}.  Such contamination is a major concern
at present, with more than 20 interactions on average at each LHC bunch crossing.  This number is 
expected to grow even larger in future runs.  Various techniques~\cite{Cacciari:2008gn,Soyez:2012hv} allow one to estimate and subtract pile-up contributions.  Approaches in which jet-substructure
analyses and
searches are done in a way which is inherently less susceptible to such contamination would offer desirable
alternatives to contamination subtraction.

Two main classes of alternative approaches have emerged: `filtering'~\cite{Butterworth:2008iy} (see also~\cite{Krohn:2009th}) and `template overlap'~\cite{Almeida:2010pa}.  In
the former, a measured jet is declustered and its soft components are removed.  This leaves only its hard
components to be reclustered into the `filtered' reclustered jet.  In the second approach
measured jets are not manipulated, and are instead compared to a set of templates built according to
a chosen (computed) fixed-order distribution of signal jets.  The comparison makes use of an `overlap
function' which evaluates the degree of overlap between each measured jet and the set of templates.
The reader will find a discussion of jet-substructure observables and experimental applications in 
Refs.~\cite{JetSubstructuresRefs1,JetSubstructuresRefs2,JetSubstructuresRefs3,JetSubstructuresRefs4}.

For both alternative approaches, it would be useful to study
distributions of the core (hard) component of jets.  This is
relatively straightforward for the signals, but much more challenging
for the QCD background.
The semi-analytic calculations we pursue here are a first step in this direction, as our results provide  a semi-analytic understanding of the kinematical
distributions of the hard component of massive jets with non-trivial
three-body kinematics.  For this purpose we also compare the result of
our full $1\to3$ calculation with a similar calculation employing an
iterated $1\to2$ collinear splitting function in the approximation to
the matrix element.  We also discuss how various scale choices affect
our result.  Finally, we compare our results with parton-shower
results, both with and without matching to tree-level matrix elements.
We cross-check our calculations with a simple analytic expression for
the planar-flow distribution in the small-planar-flow limit.

The paper is organized as follows.  In sections \ref{PlanarFlowSection} and \ref{NarrowJetsSection}, we
define the planar-flow observable, introduce narrow jets and discuss
various aspects of calculations of the jet function.  In
\sect{sec:jm}, we relate the jet functions of narrow jets to collinear
splitting functions for two-body jets.
As an illustrative example we compute
the leading-order jet function for the mass distribution
 using splitting functions. 
In \sect{sec:gj} we prove that spin correlations in
the splitting function factorize in all cases that are of interest in
this article. In \sect{sec:pf} we use these methods to compute the
leading-order combined mass and planar-flow jet function of three-body
jets. For small values of the planar flow we obtain an analytic result below, while at
large planar flow we rely on numerical integration. We also compare
the jet function to one computed using an iterated $1\to2$ approximation,
discuss the behavior of the jet function at large planar flow,
and discuss the sensitivity to scale choices.
In \sect{sec:mc}, we compare the semi-analytic jet function 
to parton-shower calculations with and without matching to 
fixed-order matrix elements.  We also discuss hadronization corrections,
and corrections from terms in matrix elements beyond the collinear
approximation.  We present our conclusions in 
\sect{sec:conc}.  Six appendices furnish a variety of
technical details.

\section{Planar Flow}
\label{PlanarFlowSection}

\def\pTjet{\pT^{\rm jet}}
\def\ppi#1{p_{\perp,#1}}
\def\vpT{{\vec p}_{\rm T}}
\def\vpTjet{\vpT^{\rm jet}}
\def\vppi#1{{\vec p}_{\perp,#1}}
We are interested in two energy flow observables --- 
the jet mass and planar flow.
The jet mass squared is,
\begin{equation}
\biggl(\,\sum_{i\in\text{jet}} p_i\biggl)^2\,, 
\end{equation}
where $i$ refers here to any parton (or hadron or tower or topocluster in a more realistic experimental context) inside the jet. 
For a hadronic collider we may define the planar flow as follows.  Define first a 
$2\times2$ momentum shape tensor,
\begin{equation}
  \cI \; \equiv
  \sum_{i\in\text{jet}} \pT^i
  \begin{pmatrix}
    \Delta\eta_i^2 & \Delta\eta_i \Delta\phi_i \\
    \Delta\eta_i \Delta\phi_i & \Delta\phi_i^2
  \end{pmatrix}\,,
\end{equation}
where 
$(\Delta\eta_i,\Delta\phi_i) = (\eta_i - \eta_{\rm jet}, \phi_i - \phi_{\rm jet})$
are the pseudorapidity and azimuthal-angle difference of
each jet constituent from the jet axis. We take all constituents to be massless. This form is manifestly boost-invariant for boosts along the
beam axis.

The planar flow is then defined by
\begin{align}
  \Pf &\equiv \frac{4 \det \cI}{(\trace \cI)^2} \ed
  \label{Pf}
\end{align}
One can easily verify that $0 \leq \Pf \leq 1$, and that it vanishes
for two-body jets, receiving its leading contribution 
from three-body jets.\footnote{Notice that because $\trace \cI$ is proportional to the jet mass~\cite{Soyez:2012hv}, the planar flow is only well defined for massive jets.} The
latter property means that the planar flow is potentially useful
for distinguishing the QCD background from top jets. The
value $\Pf = 0$ arises when the partons lie on a line in the
detector plane.  In particular, it will vanish for three-body jets when a
parton becomes soft, or when two partons become collinear. Three-parton configurations that are symmetric about the jet axis
have $\Pf = 1$.  Configurations with $n$ partons symmetric under rotations
by $2\pi/n$ radians around the jet axis also have $\Pf=1$.

For the sake of convenience, in the theoretical calculations that
follow we will focus on central jets (that is, 
with very small pseudorapidity). For such
jets we can work in terms of the angles $\theta$ and $\phi$ (being the polar coordinates around the jet axis), set
$\eta=0$, and exchange $\pT$ for $p$ (the overall jet momentum).
For narrow, central jets, we can rewrite the momentum-inertia tensor
as follows,
\begin{align}
  \cI^{kl} \; &=
  \sum_{i \in \text{jet}} \!\! E_i
  \frac{\ppi{i,k}}{E_i}
  \frac{\ppi{i,l}}{E_i}
  \ecq
  k,l=1,2 \,,
  \label{Iw}
\end{align}
where $\ppi{i,k}$ is the $k$th component of the transverse momentum of constituent $i$
with respect to the jet axis.

As we review in Sect.~\ref{sec:jm}, the leading-log behavior of the jet function for the jet mass is given by
\begin{gather}
  J_f(m^2;\pT, R) \simeq \frac{\as \tilde{c}_f}{\pi} \frac{1}{m^2}
  \log \left( \frac{\pT^2 R^2}{m^2} \right)
  \ec
\end{gather}
where in this case $\tilde{c}_g = C_A=3$ and $\tilde{c}_q = C_F=4/3$,
and the jet radius $R = \sqrt{\Delta \eta^2 + \Delta \phi^2}$. In this article we compute the mass and planar-flow jet function at leading order in $\alpha_s$, for narrow QCD jets. In the limit of small $\Pf$ we will obtain an analytic result for the leading-log behavior of the jet function. For small ${m}/{\pT R}$ it is given by the expression
\begin{gather}
  J_f(m^2,\Pf;\pT; R) =
  \frac{\as^2 c_f}{\pi^2}
  \cdot
  \frac{1}{m^2} \log\left(\frac{\pT^2 R^2}{m^2}\right)
  \cdot
  \frac{1}{\Pf} \log \left( \frac{1}{\Pf} \right)
  \ec \label{Jfanal} \\
  c_g = 2 C_A^2 \ecq
  c_q = C_F (C_A + C_F)
  \ed
  \label{c}
\end{gather}

Away from the limit of small planar flow, we will compute the planar-flow jet function semi-analytically, using numerical integration to
obtain the final result. We will show that there is a physically
interesting regime of parameters, with the jet mass near the top-quark mass
and with $0.4 \lesssim \Pf \lesssim 0.95$, in which our result has
rough agreement with the parton-shower simulations. (As shown below, the parton-shower
results of different tools do not fully agree with each other, which however is not the focus of this study.)
We expect our results to be useful for understanding how to refine
methods to distinguish highly-boosted top jets from the QCD
background at the LHC.

The planar-flow distribution was measured by CDF~\cite{Alon:2011} and ATLAS~\cite{Aad:2012meb}, but with large statistical uncertainties, and using too large a cone size and too big a mass-to-momentum ratio to be compared
with our results. At the LHC, accurate measurements
of planar-flow distributions are difficult due to pile-up effects; but we may expect them to improve significantly over time.  
In principle one can `refine' jets  in a controlled manner (by applying filtering; by using the template overlap method and then looking at the parton distribution of the peak templates; 
by looking at events with a small number of vertices; or by using other methods for pileup subtraction), and thereby isolate the hard part of the measured jet in order to compare with theoretical predictions even in the presence 
of incoherent soft radiation.

\section{Narrow Massive Jets}
\label{NarrowJetsSection}

Let us consider jets at the parton level. If we take a jet to be
narrow ($R^2 \ll 1$), the partons will be approximately collinear. As we
review in the next section, when massless partons become collinear the
QCD cross section factorizes partially, and we can write
(schematically for now),
\begin{align}
  \left. \frac{d\sigma_{2 \to JX}(R)}{d\pT \, d\eta \, d\cO}
  \right|_{\cO=\cO_0}
  &\simeq
  \sum_f \frac{d\sigma_{2 \to fX}(R)}{d\pT \, d\eta} \cdot
  J_f(\cO_0; \pT, \eta; R)
  \ed
  \label{fact}
\end{align}
In the cross section on the right-hand side, the partons making up the
jet are replaced by a parent parton of type $f$, which can be either a
gluon or one of the massless quarks. The jet substructure is encoded
in the jet function $J_f$, which has a simple physical interpretation:
it is the probability distribution for the parent parton of type $f$
to evolve into a jet of size $R$ that has $\cO=\cO_0$. Accordingly, its
integral is normalized to unity,
 $\int \! d\cO J_f(\cO) = 1$. Throughout our study we will
be agnostic about the specific jet algorithm used in the analysis, and
will assume only that it produces approximately circular jets with
radius $R$ in the $\eta$--$\phi$ plane.

As we review in the next section, 
at fixed order in $\alpha_s$ the jet function can
be computed from splitting functions~\cite{Altarelli:1977zs}, 
universal functions that govern the behavior of the squared matrix
element in the collinear limit. In this limit, the squared matrix element
factors into a product of a splitting function and a squared matrix
element with lower multiplicity.
In general, the factorization is not
complete, due to the dependence of the splitting functions on the spin
of the parent parton. We will show, however, that for all energy-flow
observables the spin dependence does factorize.

The fixed-order splitting function is singular in the limit where
partons become soft or collinear. In Eq.~\eqref{fact} this singularity
appears as a divergence of the jet function in the infrared limit of the
observable $\cO$ (for example taking $m \to 0$ in a two-body jet).
Resumming higher-order (perturbative) corrections cures the
divergence, and replaces it with a peak at a finite value of
$\cO$ (for reviews see for example
 Refs.~\cite{Sterman:2004pd,Dremin:2005kn,Han:2005mu,Sterman:1995fz}). A
fixed-order calculation is accordingly unreliable when we get close to
the infrared limit. This problem can be avoided by considering only values
of the observable $\cO$ that are far away from the infrared limit, compared
with the peak position in its distribution.  In particular, we will
always take the jet mass to be `large enough' in this sense.  The peak
in the jet-mass distribution, $m_{\rm peak}$, is roughly near its average at $m_{\rm ave} = \alpha_s
\pT R$~\cite{Ellis:2007ib}, and we will take our jet mass to be much larger, $m \gg m_{\rm
  peak}$, where we expect the fixed-order perturbative calculation to be reliable.

Beyond higher-order perturbative corrections requiring resummation,
 there are also non-perturbative corrections (in the
form of hadronization), which become important in the infrared
and tend to smear the jet function. If some of the fixed-order partons have transverse momentum relative to the jet that is small compared to the
characteristic transverse momentum of the smearing effect, the final
jet function will be dominated by the latter effects. Keeping
the observable $\cO$ away from its infrared limit avoids this problem as well. On
the other hand, in order for the collinear approximation 
to hold, we cannot stray too far from the infrared limit. Finally, in order
for the approximated distribution (in the collinear limit and at fixed
order in $\alpha_s$) to be valid, we should not get too close to
kinematic boundaries.

For our approximation to be reliable,
we need a range of values for $\cO$ that
obeys these constraints. Applying the constraints to the jet-mass
observable, and requiring that we have a non-empty range of validity
for the approximation, necessitates considering jets with sufficiently
high
$\pT$. We therefore consider only highly-boosted jets. For a general
observable, the existence of such a range of validity is less
clear. We will show later that to reasonable accuracy,
a non-trivial range of validity does
indeed exist
for the mass and planar-flow jet functions
with collision  parameters typical of the LHC.

\section{QCD Jet-Mass Distribution in the Collinear Limit}
\label{sec:jm}

In this section we compute the jet-mass distribution for massless QCD in the collinear limit using the splitting function \cite{Altarelli:1977zs}, for both quark and gluon jets. Consider again the hadronic collision,
\begin{align}
  H_A(q_A) + H_B(q_B) \to J(\cO;\pT,\eta;R) + X \ec
\end{align}
where $H_A,H_B$ are the initial hadrons with momenta $q_A,q_B$, and $J$ is a jet of cone size $R$ and given $\pT$ and $\eta$. The jet is further characterized by an energy-flow observable $\cO$. For simplicity, in this section we will take $\cO$ to be the jet mass squared, $m^2$; in the next section we will consider also the planar flow.

The factorized cross section is given by,
\begin{align}
  \frac{d\sigma_{H_A H_B \to J X}(R)}{d\pT \, d\eta \, dm^2} &=
  \sum_{f_A,f_B} \int \! dx_A \, dx_B \,
  \phi_{f_A}(x_A) \phi_{f_B}(x_B)
  \frac{d\sigma_{f_A f_B\to JX}}{d\pT \, d\eta \, dm^2}
  (x_A,x_B,\pT,\eta,m;R) \ec
  \label{hadCS}
\end{align}
where $\phi_{f_A},\phi_{f_B}$ are the parton distribution functions.  For narrow jets $d\sigma_{f_A f_B\to JX}$ further factorizes at leading order into a jet function, times a cross section in which the jet is replaced by a single parent parton,
\begin{align}
  \frac{d\sigma_{f_A f_B\to JX}}{d\pT \, d\eta \, dm^2}
  (x_A,x_B,\pT,\eta,m;R) &\simeq
  \sum_f
  \frac{d\sigma_{f_A f_B\to fX}}{d\pT \, d\eta}
  (x_A,x_B,\pT,\eta)
  \, J_f(m;\pT,\eta;R) \ed
  \label{csfact}
\end{align}
The sum is over the type $f$ of the parent parton, which can be a gluon or a quark with a specific flavor.
The relation \eqref{csfact} is due to the factorization of the QCD matrix element in the limit where two or more partons become collinear. The universal function in the factorization is proportional to the splitting function. 
We therefore seek to express the jet function in terms of the splitting function.

\subsection{Quark Jets}
\label{sec:gq}

Let us first consider the case where the parent parton is a quark.  This case is simpler because the quark splitting 
function does not depend on the helicity of the parent parton. Here the cross section factorizes completely in the 
collinear limit. Consider the matrix element for the scattering of two into $n$ massless QCD partons,
\begin{align}
  \cM_{2 \to n} \equiv
  \cM_{f_1,\dots,f_{n+2}}^{c_1,\dots,c_{n+2};s_1,\dots,s_{n+2}}
    (p_1,\dots,p_{n+2}) \ed
\end{align}
Here $p_i$ denote parton momenta; the $f_i$, parton types; the $c_i$, their colors; and the $s_i$,
their helicities. Outgoing particles are indexed by $i=1,\dots,n$, and incoming particles by $i=n+1,n+2$.
Define the abbreviation $|\cM_{2 \to n}|^2 \equiv |\cM_{f_1,\dots,f_{n+2}}(p_1,\dots,p_{n+2})|^2$ for
the squared matrix element summed over color and helicities (averaged in the case of incoming particles --- we
will leave this distinction implicit from now on).

Consider the limit in which two outgoing partons, say $p_1$ and $p_2$, become collinear. The leading contribution to $|\cM_{2 \to n}|^2$ in this limit is from diagrams in which the two outgoing partons originate from a single parent parton with momentum $p=p_1+p_2$, and with a type $f$ that is uniquely determined by the splitting process $f \to f_1 \,f_2$. In the collinear limit the parent goes on shell, leading to a $1/p^2$ singularity. 
When the parent is a quark, the squared matrix element factorizes as we approach the limit,
\begin{align}
  |\cM_{f_1,f_2,f_3,\dots,f_{n+2}}(p_1,p_2,p_3,\dots,p_{n+2})|^2 &\simeq
  |\cM_{f,f_3,\dots,f_{n+2}}(p,p_3,\dots,p_{n+2})|^2 \cdot
  \frac{8 \pi \as}{s_{12}}
  P_{f_1 f_2}(p_1,p_2) \ec
  \label{Msqrfact}
\end{align}
where $s_{12} \equiv (p_1+p_2)^2$, and $P_{f_1 f_2}$ is the spin-averaged splitting function \cite{Altarelli:1977zs,Catani:1998nv}, given in appendix~\ref{sp12-func}. In the squared matrix element on the 
right-hand side, the two collinear partons are replaced by their parent parton. For a gluon jet,
the splitting function depends on the helicity of the parent parton, and the factorization is not as simple as in Eq.~\eqref{Msqrfact}; we consider this case in the next subsection.

The fixed-order differential cross section is given in terms of the squared matrix element~\cite{Peskin:1995ev},
\begin{align}
  d\hat{\sigma}_{2\to n} &=
  \frac{1}{8E_{n+1} E_{n+2}}
  \prod_{i=1}^n \left[ \frac{1}{2 E_i} \frac{d^3 p_i}{(2\pi)^3} \right]
  |\cM_{2 \to n}|^2 \;
  \delta^4(p_{\mathrm{f,tot}} - p_{\mathrm{i,tot}}) \ed
  \label{dsigma}
\end{align}
Using Eq.~\eqref{Msqrfact} and making a change of variables $(\vec{p}_1,\vec{p}_2) \to (\vec{p}_1,\vec{p}=\vec{p}_1+\vec{p}_2)$ it is easy to see that near the collinear limit,
\begin{align}
  d\hat{\sigma}_{2\to n} \simeq
  d\hat{\sigma}_{2\to n-1} \times
  \left.
  \frac{4\pi\as}{s_{12}}
  \frac{E}{E_1 E_2} \frac{d^3p_1}{(2\pi)^3}
  P_{f_1 f_2}(p_1,p_2)
  \right|_{\vec{p}_2=\vec{p}-\vec{p}_1}
  \ec
  \label{dsigmafact}
\end{align}
where $d\hat{\sigma}_{2\to n-1}$ is defined in terms of the matrix element on the right-hand side 
of Eq.~\eqref{Msqrfact}. Comparing with the postulated relation $\eqref{csfact}$, we may now write down 
the jet function,
\begin{align}
  J_f(m^2;\vec{p};R) =
  \frac{\as E}{2\pi^2 m^2}
  \sum_{f \to f_1 f_2}
  \frac{1}{S_{f_1 f_2}}
  &\int \! \frac{d^3p_1 d^3p_2}{E_1 E_2} \, \delta^3(\vec{p}_1+\vec{p}_2-\vec{p}) \,
  P_{f_1 f_2}(p_1,p_2) 
  \notag \\ &\quad
\times \delta(m^2(p_1,p_2) - m^2) \;
  \Theta(R - \theta_1) \; \Theta(R - \theta_2)
  \ed \label{J}
\end{align}
Here, $\theta_i$ are the angles of the momenta $\vec{p}_i$ with respect to the jet axis $\vec{p}$, and the step functions $\Theta(R-\theta_i)$ are put in by hand to enforce\footnote{This procedure is expected to be compatible,
up to higher-order corrections in $R$, with any jet algorithm that produces approximately circular jets~\cite{Almeida:2008yp}.} a cone of size $R$. The sum is over allowed splitting processes, and the symmetry factor $S_{f_1 f_2}$ corrects the over-counting of identical parton configurations in Eq.~\eqref{J}: 
it is $2$ when $f_1=f_2$, and $1$ when they are different. We stress that Eq.~\eqref{J} is valid only to leading order in $\alpha_s$ at the partonic level. The full jet function receives corrections 
at higher orders in perturbation theory (some of which require resummation) as well as
from non-perturbative effects.

It is now a straightforward exercise to substitute the quark splitting function into Eq.~\eqref{J} and compute the 
jet function, assuming $R \ll 1$.
The quark splitting function is~\cite{Altarelli:1977zs,Catani:1998nv}
\begin{align}\label{SFq}
  P_{q \to gq}(z) = C_F \, \frac{1+(1-z)^2}{z}\,,
\end{align}
where $C_F=(N_c^2-1)/2N_c$, and $z=E_g/E$ is the emitted gluon's energy fraction, in our approximation. 
 Using Eq.~\eqref{SFq} and solving the integral in Eq.~\eqref{J}, we find the leading-log expression for the quark jet function, valid for $m \ll pR$
 \begin{align}
  J_q(m^2;p;R) &\simeq \frac{C_F \as}{\pi} \frac{1}{m^2}
  \log \left( \frac{p^2 R^2}{m^2} \right)
  \ed \label{Jq}
\end{align}
This result was derived in Ref.~\cite{Almeida:2008tp} using slightly different terminology but a similar limit.
As we explain in detail in appendix~\ref{sec:splitmaspf}, 
the form of~Eq.~\eqref{Jq} can be alternatively obtained by rewriting the mass as a function of the emission angle and $z$, and replacing $z$ with $m$ as the integration variable. 
This makes explicit the overall $1/m^2$ dependence, and the integration over the angle within the allowed kinematical boundaries further leads to the log in~Eq.~\eqref{Jq}.

When $m \to 0$, the jet function \eqref{Jq} diverges.  This is an infrared divergence, resulting from 
partons becoming soft and collinear. In this limit higher order contributions in $\as$ become important, and after resumming them the singularity is exponentially suppressed. Thus, the full jet function vanishes in the massless jet limit (see for example 
Refs.~\cite{Sterman:2004pd,Dremin:2005kn,Han:2005mu,Sterman:1995fz}); and because it decays at large mass, a peak of the jet mass is expected to arise at low jet mass. The result in Eq.~\eqref{Jq} is therefore 
only reliable when $m_{\rm peak} \ll m \ll p R$. The divergence in Eq.~\eqref{Jq} also renders the distribution 
non-normalizable. Nevertheless, in the regime where our approximation is valid we expect our result to match the 
full jet function including its overall normalization. In other words, in this regime it gives the (leading) probability 
distribution of a quark to evolve into a radius-$R$ jet with mass $m$~\cite{Almeida:2008tp,Almeida:2008yp}.
This behavior was verified experimentally, at least qualitatively, in Ref.~\cite{Aaltonen:2011pg}.

\subsection{Gluon Jets}
\label{sec:gj}

We now turn to the case of gluon jets.  Unlike the quark case, the gluon splitting functions have a non-trivial dependence 
on the helicity of the parent parton.  As a result, the collinear factorization of the squared matrix element is incomplete.
For the computation of the jet-mass distribution the incomplete factorization does not alter the leading-log result: in this case, the spin correlations are nonsingular.
However, we would like to go beyond this and establish a general result that will be useful in the rest of our study.
The spin correlations always factorize when one considers distributions of energy-flow observables defined for a single jet. We will prove this for the $1\to2$ splitting function, and the proof can be easily generalized to the $1\to3$ splitting function.

In the presence of spin correlations, the collinear factorization \eqref{Msqrfact} of the squared matrix element is no longer correct, and instead we have \cite{Catani:1998nv},
\begin{align}
  |\cM_{f_1,f_2,f_3,\dots,f_{n+2}}(p_1,p_2,p_3,\dots,p_{n+2})|^2 &\simeq
  \frac{8 \pi \as}{s_{12}}\sum_{ss'} \cT_{f,f_3,\dots,f_{n+2}}^{ss'}
  (p,p_3,\dots,p_{n+2})
  \hat{P}_{f_1 f_2}^{ss'}(p_1,p_2) \ec
  \label{Mfact2}
\end{align}
where now $\hat{P}$ is the helicity-dependent splitting function, and $\cT$ is defined in terms of the matrix element $\cM_{2 \to n-1}$ by
\begin{align}
  \cT_{f_1,\dots,f_{n+1}}^{s_1s_1'}(p_1,\dots,p_{n+1}) &\equiv \notag\\ &\quad
    \!\!\!\!\!\!\!\!\!\!\!\!\!\!\!\!\!\!\!\!\!\!\!\!\!\!\!\!\!\!\!\!\!\!\!\!
  \sum_{
    \begin{smallmatrix}
    s_2,\dots,s_{n+1} \\ c_1,c_2,\dots,c_{n+1}
    \end{smallmatrix}
  }
  \cM_{f_1,\dots,f_{n+1}}^{c_1,\dots,c_{n+1} ; s_1,\dots,s_{n+1}}
    (p_1,\dots,p_{n+1})
  \left[
  \cM_{f_1,\dots,f_{n+1}}^{c_1,\dots,c_{n+1} ; s_1',\dots,s_{n+1}}
    (p_1,\dots,p_{n+1})
  \right]^\dagger \ed
  \label{T}
\end{align}
$\cT$ is essentially the matrix element $\cM_{2 \to n-1}$, squared, except that there is no sum over the helicity
of the jet's parent parton.

The differential cross section no longer factorizes as it did in Eq.~\eqref{dsigmafact}.
 However, in the jet-function definition the phase-space integral includes an azimuthal integral around the jet axis. For example, changing the integration variables in Eq.~\eqref{J} to polar coordinates $(p_i,\theta_i,\phi_i),\,i=1,2$ (relative to the jet axis $\vec{p}$), the integration over rotations around the jet axis is described by the variable $\phi\equiv\phi_1$, keeping $(\phi_2-\phi_1)$ fixed.
Therefore, for any observable that is invariant under rotations around the jet axis
the integral picks out the part of the splitting function that is invariant under such rotations, which is precisely the spin-averaged splitting function. This includes all energy-flow observables defined in terms of a single jet, because for such observables there is no preferred direction in the detector plane that can break the rotational symmetry.
We may therefore replace $\hat{P}^{ss'} \to \delta^{ss'} P$, where $P$ is the spin-averaged splitting function. Noting that $\cT^{ss'} \delta^{ss'} = |M_{2 \to n-1}|^2$, the rest of the computation follows through as in the previous section, and we conclude that the jet function for gluon jets is given by Eq.~\eqref{J}, just as for quark jets.

We conclude this section by computing the leading-log part of the gluon jet-mass function. 
For $R^2\ll 1$ and $m^2 \ll (p R)^2$, the leading contribution comes only from the $g \to gg$ splitting function,
\begin{align}
  P_{gg}(z) &= 2 C_A \left[ \frac{z}{1-z} + \frac{1-z}{z} + z(1-z) \right] \ec
\end{align}
where $C_A = N_c$, and we find
\begin{align}
  J_g(m^2;p;R) &\simeq \frac{C_A \as}{\pi} \frac{1}{m^2}
  \log \left( \frac{p^2 R^2}{m^2} \right)
  \ec \label{Jg}
\end{align}
again in agreement with Ref.~\cite{Almeida:2008tp}.

\section{Planar-Flow Jet Function}
\label{sec:pf}

In this section we compute the planar-flow jet function $J_f(m^2,\Pf;\pT;R)$, which receives its leading contribution from three-body jets. This jet function factorizes from the rest of the cross section when we take the ``triple'' collinear limit, in which three partons become collinear simultaneously. The limit is analogous to the (``double'') collinear limit that we studied in the last section.

We consider $2 \to n$ scattering with matrix element $\cM$, where three outgoing massless partons with 
four-momenta $p_1, p_2, p_3$ are collinear. The parton energies are denoted by $E_i$. The jet made out of these three partons has energy $E = \sum_{i=1}^3 E_i$, momentum $p$ and mass $m$. Finally, we define the energy fractions $z_i \equiv E_i/E$, and also,
\begin{align}
  s_{ij} &\equiv (p_i+p_j)^2 = 2 p_i \cdot p_j \ecq i,j=1,2,3 \ec \nonumber \\
  s_{123} &\equiv (p_1+p_2+p_3)^2 = m^2 \ed
\end{align}

In the collinear limit, the squared matrix element of the scattering process factorizes as follows~\cite{Catani:1998nv},
\begin{align}
  |\cM_{f_1,f_2,f_3,\dots,f_{n+2}}(p_1,p_2,p_3,\dots,p_{n+2})|^2 \simeq
  \frac{64 \pi^2 \as^2}{s_{123}^2}
  \cT^{ss'}_{f,f_4,\dots,f_{n+2}}(p,p_4,\dots,p_{n+2})
  \hat{P}^{ss'}_{f_1 f_2 f_3}(p_1,p_2,p_3) \ed
  \label{M3}
\end{align}
Here, $\hat{ P}^{ss'}_{f_1 f_2 f_3}$ is the one-to-three splitting function \cite{Campbell:1997hg,Catani:1998nv} for the splitting $f\to f_1f_2f_3$; the type $f$ of the parent parton is determined by flavor conservation. The splitting function depends on the outgoing momenta and each parton $f_i$ is associated to a momentum $p_i$.
As in Sect.~\ref{sec:gj}, performing the azimuthal-angle integrals will make spin correlations disappear
in the jet function. We may thus replace $\cT^{ss'} \hat{P}^{ss'}_{f_1f_2f_3}$ by $|M_{2 \to n-1}|^2  P_{f_1f_2f_3}$, where $P_{f_1f_2f_3}$ is the spin-averaged splitting function. In the jet function we will need to sum over all possible final partons and, as before, take into account the symmetry factor that compensates for an over-counting of 
identical-parton configurations. Let us define then
\begin{align}
{P}_f\equiv\sum_{f\to f_1f_2f_3}\frac{1}{S_{f_1f_2f_3}}P_{f_1f_2f_3} \ec
\label{Ptf}
\end{align}
where $S_{f_1f_2f_3}$ equals $n!$ for a parton configuration that includes $n$ identical partons. Accordingly,
\begin{align}
  {P}_g &=
  \frac{1}{3!} P_{ggg} + N_f P_{gq\bar{q}} \ec \label{Ptg} \\
  {P}_q &=
  \frac{1}{2!} P_{ggq} +
  \frac{1}{2!} P_{qq\bar{q}} +
  (N_f-1) P_{\bar Q Qq} \ec
  \label{Ptq}
\end{align}
where $Q$ denotes a quark with a different flavor than that of the parent $q$. The spin-averaged splitting functions $P_{f_1f_2f_3}$ are given in appendix~\ref{sp-func}.

The jet function is given by a straightforward generalization of Eq.~\eqref{J},
\begin{align}
  J_f(m^2,\Pf;\vec{p};R) &\simeq
  \frac{\as(m) \as(\mu) E}{4\pi^4 m^4}
  \int \!
  \prod_{i=1}^{3} \left[ \frac{d^3p_i}{E_i} \Theta(R - \theta_i) \right]
  {P}_f(\vec{p}_i)
  \, \delta\! \left( m^2(\vec{p}_i) - m^2 \right) \,
  \notag\\ &\quad \times
  \delta\!\left( \Pf(\vec{p}_i) - \Pf \right) \,
  \delta^3\!\left( \vec{p} - \sum \vec{p}_i \right)
  \ed
  \label{Jf1}
\end{align}
As discussed in detail further below (see subsection~\ref{subsec:runScale}), the two strong-coupling factors are evaluated
at different scales\footnote{For the
  sake of brevity we allow ourselves to be sloppy in our notation: generally $\mu$ in Eq.~\eqref{Jf1} depends on
  the parton configuration in the integrand, so $\alpha_s(\mu)$ should really appear inside the integral.}: we take the
first scale to be $m$, corresponding to the first $1 \to 2$ splitting, and
we choose the second scale to approximate that of the second splitting 
according to an ordering scheme defined below
(obviously for ``Mercedes''-like configurations the ordering does not make a difference). 

An equivalent definition of  the jet function~\eqref{Jf1} is provided by averaging the splitting function over parton
permutations,
\begin{align}
P_f \;\longrightarrow\; \tilde P_f\equiv \frac{1}{6}\sum_{\sigma\in S_3}P_f(p_{\sigma(1)},p_{\sigma(2)},p_{\sigma(3)})~,
\label{Pftilde}
\end{align}
while restricting the integration domain to
\begin{align}
\int_{-\infty}^{\infty} d^3p_1d^3p_2d^3p_3 \;\longrightarrow\; 3 \int_{E_1<E_2,E_3} d^3p_1d^3p_2d^3p_3~.
\label{parD2}
\end{align}
The motivation for this alternative formulation will become clear below, but the basic idea will be to keep one parton 
(say, $\vec p_3$) away from its soft limit $E_3\to 0$. We can do this without loss of generality as our jet is assumed to be of a large mass.

Let us now integrate over $\vec{p}_3$ in Eq.~\eqref{Jf1} by using momentum conservation. In the remaining integral let us switch to spherical coordinates relative to the jet axis $\vec{p}$, $(p_i,\theta_i,\phi_i)$, $i=1,2$.

Next, we extract the leading-order expressions for the integrand
in the narrow-jet expansion $\theta_i<R \ll 1$.
The resulting expressions appear in appendix~\ref{kinematics}.  Using
these, as well as the explicit splitting functions, one may easily check that the
integrand in Eq.~\eqref{Jf1} depends on the azimuthal angles $\phi_i$ only through
the combination $\cos(\phi_1-\phi_2)$.  We make the change of variables
$\int_0^{2\pi} d\phi_1 d\phi_2 f(\cos(\phi_1-\phi_2)) = 4\pi \int_0^\pi d\phi
f(\cos(\phi))$.

We can now write the jet function as follows,
\begin{align}
  J_f &\simeq
  \frac{3 \as^2 p}{\pi^3 m^4}
  \int_0^R \! d\theta_1 d\theta_2 \,
  \int_0^\pi \! d\phi
  \int_\cC \! dp_1 \, dp_2 \,
  \frac{p_1 p_2 \theta_1 \theta_2}{p_3}
  \tilde{P}_{f}(\vec{p}_i) \,
  \delta\!\left( m^2(\vec{p}_i) - m^2 \right)
  \delta\!\left( \Pf(\vec{p}_i) - \Pf \right)\,,
  \label{Jf}
\end{align}
where the domain of integration is,
\begin{align}
  \cC &\equiv
  \left\{
    (p_1,p_2)
    \ \left|\ 0 \le p_1\le p_2,p_3 \ec\; \theta_3\le R \right.
  \right\} \ed
  \label{C}
\end{align}
Here, $p_3$ and $\theta_3$ are given (in terms of our integration variables) by
eqs.~\eqref{p32} and~\eqref{theta3} respectively. It can be easily seen that the resulting collinear expansion of $p_3$, Eq.~\eqref{p32}, breaks down for small $p_3$. In other words, the soft limit of $p_3$ and the collinear limit do not commute. Therefore, one cannot use the expansion \eqref{p32} in the vicinity of small $p_3$. We avoid this complication by formulating the jet function in a way that avoids that part of the integration region, as in Eq.~\eqref{parD2}.

Using Eq.~\eqref{theta3}, the inequality $\theta_3 \le R$ is equivalent at leading order to
\begin{align}
  p \left( R^2 - \frac{m^2}{p^2} \right) -p_1(R^2 - \theta_1^2)  -p_2 (R^2 -
  \theta_2^2)  \ge 0  \ed
  \label{th3-cone}
\end{align}
Notice that we need $m < p R$ to have a non-vanishing domain, in accordance with the narrow jet approximation. Very close to the kinematic boundary the leading contribution will be dominated by higher order terms; consequently, we must
require $m$ to be safely away from the kinematic boundary, $m^2\ll (p R)^2$.

Before proceeding, let us discuss the expected range of validity of our theoretical computation.  Take 
$\pT \simeq p = 1~\TeV$. The collinear approximation requires that $R^2 \ll 1$; we choose $R=0.4$ for our analyses, 
the same as the smallest cone size used at the LHC for new-physics searches. As for the jet mass, we would like to stay well below the kinematic boundary at $p R$, but also well above the peak in the mass distribution (very roughly at $\alpha_s p R$), so that higher-order effects in $\alpha_s$ remain small. We choose $m \simeq 180~\GeV$, which satisfies both of
these constraints.  This choice is also well-motivated physically, as it is close to the top-quark mass.
The singularity at small $\Pf$ would be eliminated by resummation of higher powers of $\log\Pf$; we would expect
such a resummation to lead to a Sudakov-like exponential damping term, $\sim \exp\left(-\log^2 \Pf\right)$, similar to
the case of the thrust or jet-mass distribution~\cite{Catani:1992ua}. To the best of our knowledge no computation of this effect has been carried out to date.  In any case the focus of our present investigation is on
massive jets with sizeable planar flow where higher-order corrections, resummed or not,
have a subdominant impact. Qualitatively we expect the $\Pf$ distribution to be similar to that of the jet mass, namely, vanishing for $\Pf=0$, peaking at a small value of $\Pf$, and falling gradually beyond that point.
As we shall see in Sect.~\ref{sec:mc}, the jet parameters chosen above lead to a Sudakov peak in the planar-flow distribution near $\Pf = 0.1$\,.  We will not restrict the range of planar-flow values in our discussion, but remind the
reader that physically reliable results are expected within our framework only for planar-flow values well above this peak.

\subsection{Analytic Leading-Log Behavior}
\label{LL}

In this section we compute the jet function analytically in the limit of small planar flow and at fixed jet mass. For simplicity, we will take the second running scale $\mu$ in Eq.~\eqref{Jf1} equal to the first, $m$. We will find that the leading-log result, in the limit of small $\Pf$ and small $m/pR $, is given by Eq.~\eqref{Jfanal}.

At $\Pf=0$, the integral in Eq.~\eqref{Jf} diverges because the splitting function is singular. The singularity arises
in regions of integration where a parton becomes soft, or two partons become
collinear (with respect to the third parton). In fact, the leading singularity in the integrand arises in the
combined soft-collinear limit, where a single parton becomes both soft and
collinear with another parton. In the splitting functions (see appendix \ref{sp-func}), the terms responsible
for this leading singularity are those proportional\footnote{One might have expected terms
proportional to $s_{ij}^{-2}$ to lead to an even higher singularity. A careful examination of these terms, though, reveals that they are in fact less singular than the other ones mentioned above.} to $(s_{ij} z_i)^{-1}$ or $(s_{ij}
s_{ik})^{-1}$.

At small (but nonzero) $\Pf$, the leading contribution to the (finite) integral will thus come from the soft-collinear regions, which are disconnected in the domain of integration. In Eq.~\eqref{Jf} there are two such regions, where parton 1 is soft, and collinear with either parton 2 or 3. Due to the symmetry of the partons, it suffices to compute the contribution from one region. We will compute the contribution from the region in which parton 1 is soft ($p_1\to 0$), and collinear with parton 2 ($\theta_{12} \to 0$).

Let us choose more convenient variables in Eq.~\eqref{Jf} to describe the collinear limit,
\begin{align}
  \theta_1 = \theta \ecq
  \theta_2 = \theta (1 + r \cos\omega) \ecq
  \phi = r \sin\omega \ed
\end{align}
To leading order in $r$, we see from Eq.~\eqref{theta12} that $\theta_{12} = \theta r$, and the collinear limit is now parameterized by $r \to 0$. 

In the soft-collinear limit ($p_1,r\to 0$) the splitting function can be written as
\begin{align}
  \tilde{P}_f = \frac{F_f(\theta;p,m)}{p_1^2 r^2} + \cdots \ec
\end{align}
where the dots include less singular terms; this is the leading singularity to which we alluded above. The functions $F_f$ are given by
\begin{align}
  F_g &= 2 \, \frac{
  C_F T_R N_f m^2 p^2 \theta^2 \left(m^4+p^4 \theta^4\right)
  + C_A^2 \left(m^4+m^2 p^2 \theta^2+p^4 \theta^4\right)^2}
  {3 p^2 \theta^4 \left(m^2+p^2 \theta^2\right)^2} \ec
  \\
  F_q &= C_F \, \frac{2 C_F m^6+(C_A+2 C_F) m^4 p^2 \theta^2
  +(2 C_A+C_F) m^2 p^4 \theta^4+2 C_A p^6 \theta^6}
  {3 p^2 \theta^4 \left(m^2+p^2 \theta^2\right)} \ed
\end{align}
As we seek the leading $\Pf\rightarrow0$ singularity, we will include only these terms in the splitting function.

Using Eqs.~\eqref{p32}, \eqref{mcol} and \eqref{pfcol}, at leading order in $p_1,r$ and in the collinear approximation we have
\begin{align}
  m^2 &= \frac{p^2 p_2 \theta^2}{p-p_2} \ecq
  \Pf = \frac{4 p^3 p_1 p_2 \theta^4 r^2 \sin^2 \omega}{m^4 (p-p_2)} \ecq
  p_3 = p - p_2 \ed
\end{align}
The integration domain \eqref{C} becomes\footnote{As we are interested only in the region where $p_1$ is soft, $p_1 \le p_2,p_3$ is automatically satisfied at a fixed mass.}
\begin{align}
  \cC &\simeq
  \left\{
    (p_1,p_2)
    \ \left|\
    p_1 \ge 0 \ec\;
    (R^2 - \theta^2) p_2
    \le p \left( R^2 - \frac{m^2}{p^2} \right)
    \right.
  \right\} \ed
\end{align}

The jet function \eqref{Jf} can now be written as
\begin{align}
  J_f &\simeq
  \frac{6 \as^2 p}{\pi^3 m^4}
  \int_0^R d\theta \int_0^{\pi} d\omega
  \int_\cC dp_1 dp_2
  \int_0^{r^{\mathrm{max}}} dr \,
  \frac{p_1 p_2 \theta^3 r}{p-p_2}
  \times \notag \\ &\quad
  \quad \delta \left[ \frac{p^2 p_2 \theta^2}{p-p_2} - m^2 \right]
  \delta \left[ \frac{4 p^3 p_1 p_2 \theta^4 r^2 \sin^2 \omega}{m^4 (p-p_2)} - \Pf \right]
  \tilde{P}_f(\vec{p}_i)
  \ec
\end{align}
with a factor of 2 from the sum over the soft-collinear regions.
The upper integration limit for $r$ is a function of the other variables whose precise form is irrelevant
to the leading behavior we are trying to compute.

In the limit of small planar flow, the dominant contribution to the jet function is from the integration region near the soft-collinear singularity.  We may therefore restrict our domain of integration to a small region around the singularity, defined by
\begin{align}
0<p_1<p_1^{\mathrm{max}} \ecq 0<r<r^{\mathrm{max}} \ec
\end{align}
where now both $r^{\mathrm{max}}$ and $p_1^{\mathrm{max}}$ are chosen constant and small. We expect that, in the small $\Pf$ limit, the jet function will not depend on these parameters. Indeed, we will see that $r^{\mathrm{max}}$ and $p_1^{\mathrm{max}}$ will drop out of the result.

We can now solve the delta functions for $p_1$ and $p_2$, 
\begin{align}
 p_1 = \frac{m^2 \Pf}{4p \theta^2 r^2\sin^2\omega} \ecq  p_2 = \frac{ m^2p}{m^2+p^2 \theta^2} \ed
\end{align}
Taking into account the Jacobian and the appropriate domain, we find,
\begin{align}
  J_f &\simeq
  \frac{6 \as^2 p^2}{\pi^3 m^2 Pf}
  \int_{m^2/p^2 R}^R d\theta  \frac{\theta^3 F_f(\theta;p,m)}{(m^2+p^2\theta^2)^2}
  \int_0^{\pi} d\omega
  \int_{\frac{m}{\theta\sin\omega}\sqrt{\frac{\Pf}{4p p_1^{\mathrm{max}}}}}^{r^{\mathrm{max}}} 
  \frac{dr}{r}
  \ec
  \label{Jf2}
\end{align}
where the lower integration limit on $r$ emerges from the previous upper integration limit on $p_1$.
Proceeding with the $r$ integration, we finally find the leading-log approximation for the jet function,
\begin{align}
  J_f \simeq \frac{A_f}{\Pf} \log \left( \frac{1}{\Pf} \right) + \cdots \ec
\label{JLL}
\end{align}
where
\begin{align}
  A_f \equiv \frac{3 \as^2 p^2}{\pi^2 m^2}
  \int_{m^2/p^2 R}^R d\theta
  \left. \frac{\theta^3 F_f \left( \theta,p,m \right)}
  {(p^2\theta^2 + m^2)^2}  \right.
  \ed
  \label{A}
\end{align}
As anticipated, the arbitrary integration limits $r^{\mathrm{max}}$ and $p_1^{\mathrm{max}}$ do not appear in the leading term. 
When we include the first correction, the jet function takes the form,
\begin{align}
  J_f \simeq \frac{A_f}{\Pf} \log \left( \frac{B_f}{\Pf} \right) + \cdots\ed
\label{JsLL}
\end{align}
Here we will not compute $B_f$, which does depend on the values of $r^{\mathrm{max}}$ and $p_1^{\mathrm{max}}$. The remaining integral in Eq.~\eqref{A} can be performed analytically, and the full result is given in appendix \ref{AfAppendix}. Here we record only the result to leading order in ${m}/{pR}$,
\begin{align}
  A_f &\simeq
  \frac{\as^2 c_f}{\pi^2}
  \frac{1}{m^2}
  \log\left(\frac{p^2 R^2}{m^2}\right) \ec
\label{AfAnalytic}
\end{align}
where $c_g = 2 C_A^2$, $c_q = C_F (C_A + C_F)$.
This is the same leading-log behavior that we found in the case of the jet-mass function.

For the relevant range of parameters as taken earlier, the contributions subleading
in ${m}/{pR}$ cannot be neglected. Taking $p=1~\TeV$,
$m=180~\GeV$, $R=0.4$, and evaluating the coupling at the jet mass
scale, we find $A_g = 0.683~\TeV^{-2}$, $A_q = 0.227~\TeV^{-2}$. In the
next section we will compare this result with a full semi-analytic evaluation of
the jet function in the collinear approximation.

In appendix~\ref{sec:splitmaspf} we show how, in parallel to the evaluation of the jet-mass distribution, the most singular term in the jet-$\Pf$ distribution can be obtained simply by iterating the $1\to2$ splitting function. The $1/\Pf$ factor arises from changing variables for the second emission-energy fraction while the $\log(\Pf)$ factor results from integrating the
angular variable within the kinematic constraints. 

\subsection{Semi-Analytic Evaluation}
\label{numeval}

Let us now return to the expression \eqref{Jf} for the jet function, and
proceed without making further assumptions. We will carry out two integrations using the delta function constraints, and compute the remaining integrals numerically to obtain the jet function. This computation is valid in the collinear approximation and for any planar flow.

\begin{figure}[H]
  \centering
  \subfigure[~Small $\Pf$]{
    \includegraphics[width=0.45\textwidth]{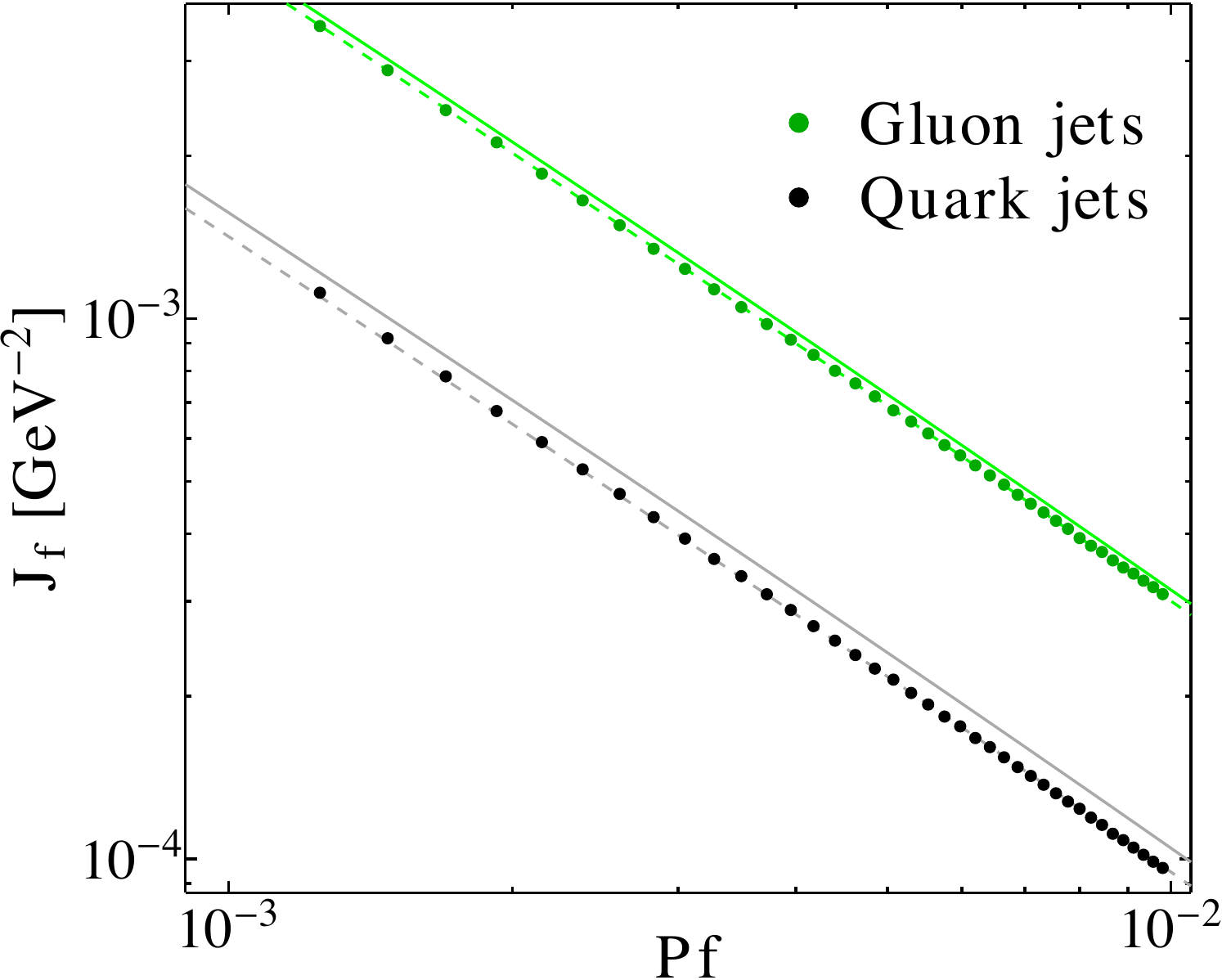}
    \label{fig:fit-small-pf}
  }
  \subfigure[~Large $\Pf$]{
    \includegraphics[width=0.45\textwidth]{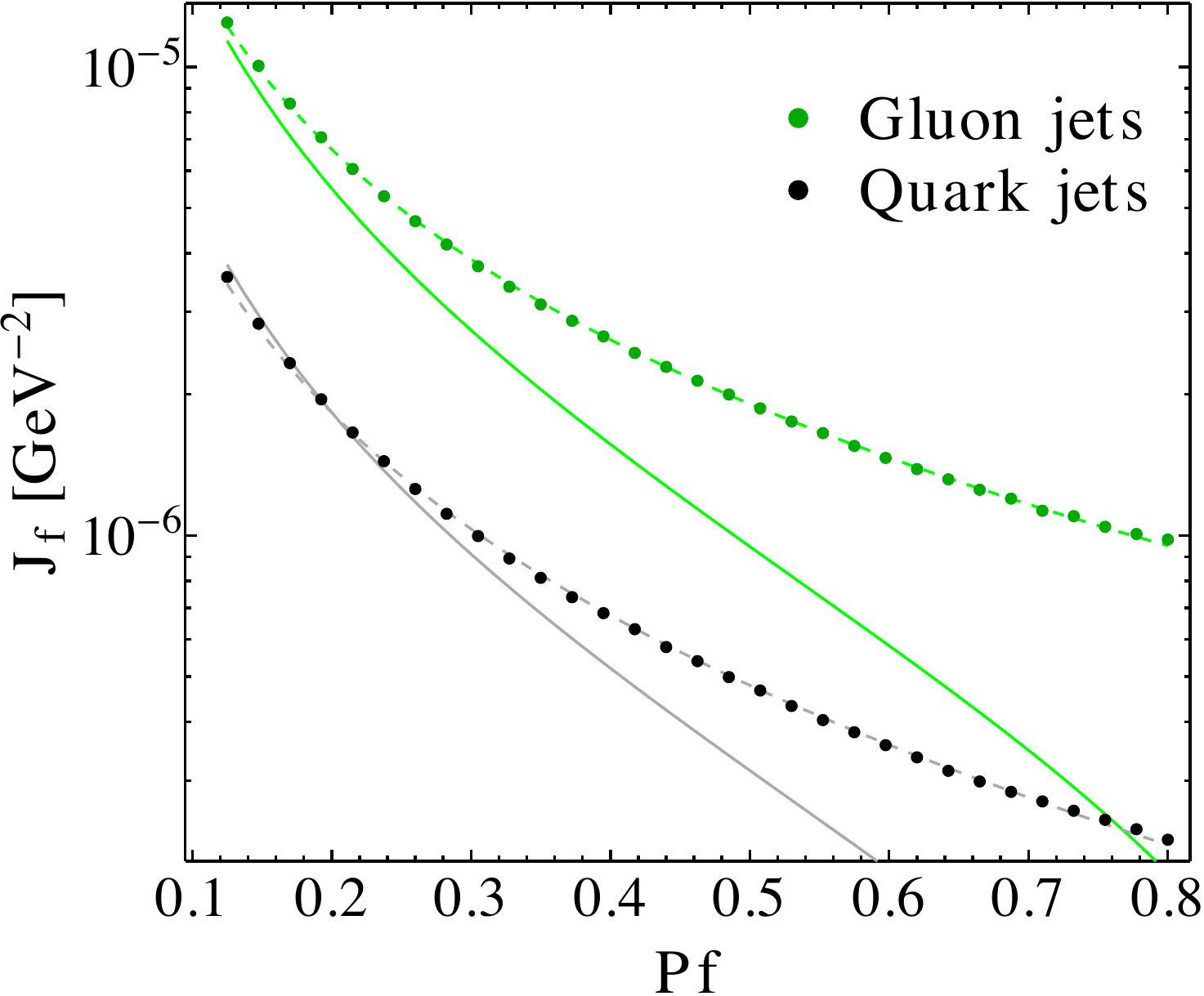}
    \label{fig:fit-large-pf}
  }
  \caption{
  Comparison of semi-analytic jet functions (dots) with the purely analytic leading-log form of Eq.~\eqref{JsLL} (solid lines) and with a fit to the leading-log form (dashed lines).  The quark functions are the lower (black) curves and the gluon ones are the upper (green) curves.  Two ranges of the planar-flow observable are shown:  \subref{fig:fit-small-pf} $10^{-3} < \Pf < 10^{-2}$
    and~\subref{fig:fit-large-pf} $0.1 < \Pf < 0.8$. 
     The curve parameters are listed in table~\ref{fit-parameters}; the purely analytic parameters are given in column 1, the fit parameters for \subref{fig:fit-small-pf}  in column 2, and the fit parameters for \subref{fig:fit-large-pf}  in column 3.  In the purely analytic form, $A_f$
is given by Eq.~\eqref{AfAnalytic} , and $B_f=1$.  In the fit to the leading-log form, $A_f$ is determined by fitting in both  \subref{fig:fit-small-pf} and \subref{fig:fit-large-pf}, while $B_f=1$ is set to 1 in (a) and is determined by fitting in (b).
The jet momentum
    is $p=1~\TeV$, with mass $m=180~\GeV$ and size $R=0.4$.
 }
  \label{fig:fit-pf}
\end{figure}

Let us perform the integration in Eq.~\eqref{Jf} over $p_1$ and $p_2$ by using the mass and planar-flow constraints. Solving these constraints, one finds two solution branches $p_i^\pm$ that should be summed over,
\begin{align}
  p_1^{(\pm)}(\theta_1,\theta_2,\phi) &=
  \frac{2 m^2 p^3 \theta_1^2 \theta_2^2 \sin^2\!\phi
  + m^4 \, p \, \Pf \, \theta_2 (\theta_2-\theta_1\cos\phi ) \pm
  \sqrt{\Delta}}{
  4 p^2 \theta_1^2 \theta_2^2 \sin^2\!\phi \left(m^2 + p^2 \theta_1^2\right)
  + m^4 \Pf \left(\theta_1^2 + \theta_2^2 -2 \theta_1 \theta_2 \cos\phi \right)}
  \ec\\
  p_2^{(\pm)}(\theta_1,\theta_2,\phi) &=
  p_1^{(\mp)}(\theta_2,\theta_1,\phi)
  \ec
\end{align}
where
\begin{align}
  \Delta &\equiv m^4 p^2 \theta_1^2 \theta_2^2 \sin^2\!\phi
  \left\{
  2 p^2 \theta_1 \theta_2
  \left[p^2 \theta_1 \theta_2 (1-\cos(2 \phi)-2 \Pf) -2 m^2 \Pf \cos\phi \right]
  -m^4 \Pf^2
  \right\}
  \ed
\end{align}
The jet function becomes
\begin{align}
  J_f &\simeq
  \frac{6 \as^2 p}{\pi^3 m^4}
  \sum_{s\in\{+,-\}}
  \int_{\cC_s'} \! d\theta_1 d\theta_2 d\phi \,
  \frac{p_1^{(s)} p_2^{(s)} \theta_1 \theta_2}{p_3}~
  |\cJ^{(s)}| ~ \tilde{P}_f(\vec{p}_i)
  \ec
  \label{final-int}
\end{align}
where $\cJ^{(s)} = \partial(p_1^{(s)},p_2^{(s)})/\partial(\Pf,m^2)$ is the Jacobian, and the new integration domain (including a reality condition for $p_i^{(\pm)}$) is
\begin{align}
  \cC'_s = \, \Bigr\{
    (\theta_1,\theta_2,\phi) \ &\left| \
    \Delta \ge 0 \ecq 0 \le p_1^{(s)} \le p_2^{(s)}, p_3\mskip -1mu\bigl( p_i^{(s)}, \theta_i, \phi \bigr) \ecq \right.
    \notag\\ &\quad
     0 \le \theta_{1,2} \le R \ecq
    0 \le \phi \le \pi \ecq     \theta_3\mskip -1mu\bigl( p_i^{(s)}, \theta_i, \phi \bigr) \le R 
  \, \Bigr\} \ed
\end{align}
We compute the integral in Eq.~\eqref{final-int} numerically.\footnote{We obtained all semi-analytic results in this paper using {\sl Mathematica\/} with adaptive Monte-Carlo integration.} For the comparison with the analytic result we take both couplings at the jet-mass scale; when comparing with the parton-shower simulations we take the first coupling at the jet-mass and the second coupling at the dipole scale (see subsection \ref{subsec:runScale} and appendix \ref{renormalization}). Fig.~\ref{fig:fit-pf} shows the results, including a fit to the predicted
leading-log behavior.  Table \ref{fit-parameters} compares the leading-log
coefficients predicted in Sect.~\ref{LL} with those extracted from the fit.
At planar-flow values below $10^{-2}$ we find good agreement with the predicted
leading-log behavior. At larger values of the planar flow the subleading behavior
becomes important, and the fit must include the subleading coefficient
$B_f$ as well. We stress that in the small-$\Pf$ region, higher-order effects
are very important, so that the planar-flow jet function obtained here
requires resummation and cannot be compared usefully with experimental
data.
Above $\Pf = 10^{-1}$ the leading coefficient $A_f$ no longer
agrees with the analytic prediction, which assumes $\Pf \ll 1$, but the jet function does match the general form
in Eq.~\eqref{JsLL}. In this
region, we must integrate the splitting functions numerically to
obtain a semi-analytic prediction.

\begin{table}
  \begin{tabular}{c||c|c|c}
    & ~Analytic~
    & ~Fit at Small $\Pf$~
    & ~Fit at Large $\Pf$~
    \\ \hline
    ~$A_g$~ & $0.683$ & $0.651$ & $0.413$
    \\ \hline
    ~$A_q$~ & $0.227$ & $0.205$ & $0.138$
    \\ \hline
    ~$B_g$~ & 1  & fixed to 1 & $5.06$
    \\ \hline
    ~$B_q$~ & 1  & fixed to 1 & $2.85$    
  \end{tabular}
  \caption{Fit coefficients for the leading-log form in Eq.~(\ref{JsLL}). The coefficients $A_f$ are in units of
  $\TeV^{-2}$, where the $B_f$ are dimensionless. The analytic result is only for the leading-order term (in small $\Pf$), so that effectively $B_f=1$.   }
  \label{fit-parameters}
\end{table}

\subsection{Comparison with Iterated $1\rightarrow2$ Splittings}
\label{1to2}

In this section we obtain a different semi-analytic approximation to
the jet function, by
approximating each $1\to 3$ splitting function by an iteration
of two $1\to 2$
splitting functions. For simplicity, we ignore spin correlations, which do not contribute at leading order in small $\Pf$.\footnote{The $1\to 3$ and $(1\to 2)^2$ splitting functions can be shown to have the same soft-collinear leading singularity.}
This corresponds to the strongly ordered limit for the two splittings.

\begin{figure}[H]
  \centering
  \subfigure[]{
    \includegraphics[width=0.45\textwidth]{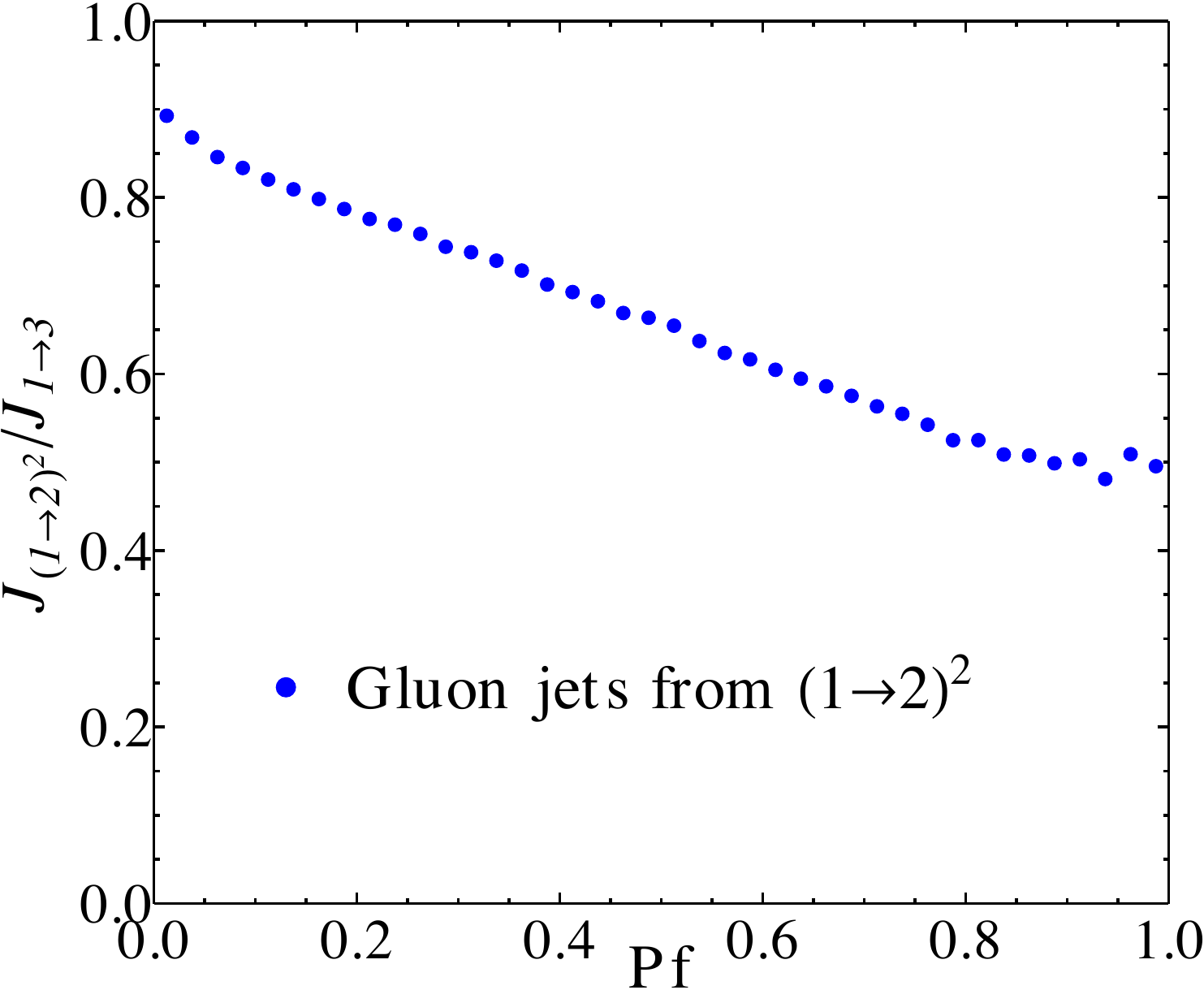}
  }
  \qquad
  \subfigure[]{
    \includegraphics[width=0.45\textwidth]{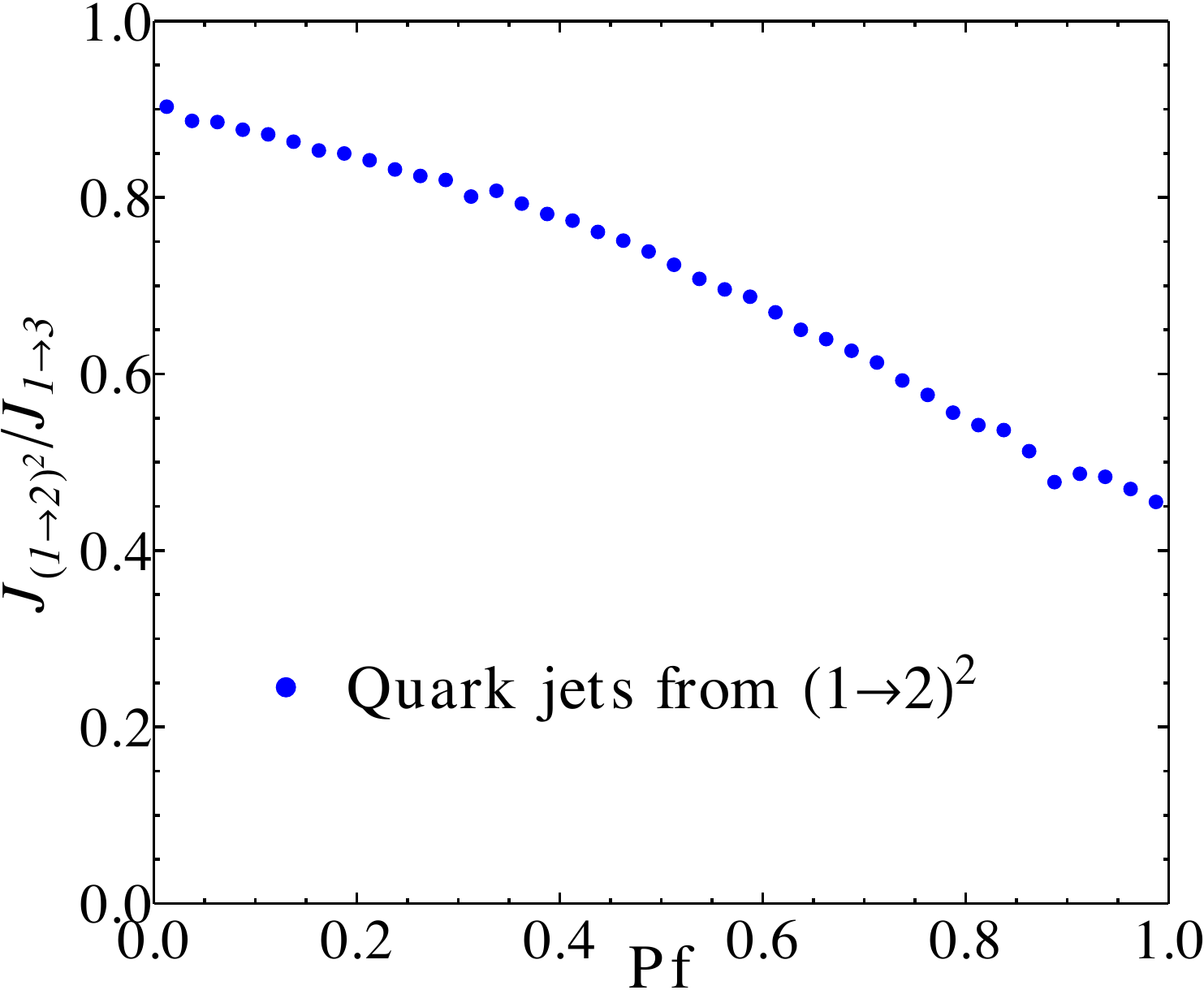}
  }
  \caption{The ratio of semi-analytic jet functions evaluated with a
    splitting kernel approximated by an iterated $1\to 2$ splittings,
    to those with the original $1\rightarrow3$ kernel.  The parameters are
    $p=1~\TeV$, $m=180~\GeV$ and $R=0.4$, and the couplings are
    evaluated at the jet-mass scale.  }
  \label{fig:1to2}
\end{figure}

Consider again the squared matrix element
\begin{align}
  |M_{2 \to n}|^2 =
  |M_{f_1 f_2 f_2 \dots f_{n+2}}(p_1,p_2,p_3,\dots,p_{n+2})|^2 \ed
\label{Msq}
\end{align}
In the limit where $p_1$ and $p_2$ are collinear, that is 
have relative transverse momentum small compared to all other parton pairs,
the matrix element~\eqref{Msq} factorizes [cf. Eq.~\eqref{Msqrfact}] as,
\begin{align}
  |M_{2\to n}|^2 \simeq
  |M_{f',f_3,\dots,f_{n+2}}(p',p_3,\dots,p_{n+2})|^2 \cdot
  \frac{8\pi \alpha_s}{s_{12}} P_{f_1 f_2}(p_1,p_2) \ec
\end{align}
where $p'$ denotes the parent parton of partons 1 and 2, namely
$p'=p_1+p_2$, and $f'$ is determined by $f_1$ and $f_2$. The splitting
functions $P_{f_1 f_2}$ are given in appendix~\ref{sp12-func}. Next, assume that $p'$ and $p_3$ are also collinear,
with relative transverse momentum small compared to that of
remaining parton pairs (but large compared to that of $p_1$ and $p_2$).
We then have the further factorization,
\begin{align}
  |M_{2\to n}|^2 \simeq
  |M_{f,f_4,\dots,f_{n+2}}(p,p_4,\dots,p_{n+2})|^2 \cdot
  \frac{64\pi^2 \alpha_s^2}{s_{12} s_{123}}
  P_{f_1 f_2}(p_1,p_2)
  P_{f' f_3}(p',p_3)
  \ec
\end{align}
where $p$ and $f$ are the jet's momentum and type,
respectively. Comparing with Eq.~\eqref{M3}, we can read off the iterated
$1\to 2$ approximation to the function ${P}_f$ defined in Eq.~\eqref{Ptf},
\begin{align}
  {P}_f^{(1\to 2)^2} =
  \frac{s_{123}}{s_{12}} \sum_{f\to f' f_3 \to f_1f_2f_3} \frac{1}{S_{f_1 f_2}}
  P_{f' f_3}(p',p_3) P_{f_1 f_2}(p_1, p_2) \ed
\label{Pt12}
\end{align}
Explicitly,
\begin{align}
  {P}_g^{(1\to 2)^2} &=
  \frac{s_{123}}{s_{12}}
  \Big[
  \frac{1}{2} P_{gg}(p',p_3) P_{gg}(p_1,p_2)
  + N_f P_{gg}(p',p_3) P_{q\bar{q}}(p_1,p_2)
  \notag \\ &\quad \qquad \;\;
  + 2 N_f P_{q\bar{q}}(p',p_3) P_{qg}(p_1,p_2)
  \Big]
  \label{Pt12g}
  \ec \\
  {P}_q^{(1\to 2)^2} &=
  \frac{s_{123}}{s_{12}}
  \Big[
  P_{qg}(p',p_3) P_{qg}(p_1,p_2)
  + \frac{1}{2} P_{gq}(p',p_3) P_{gg}(p_1,p_2)
  \notag \\ &\quad \qquad \;\;
  + N_f P_{gq}(p',p_3) P_{q\bar{q}}(p_1,p_2)
  \Big]
  \label{Pt12q}
  \ed
\end{align}
The strongly ordered approximation to the
planar-flow jet function is given by Eq.~\eqref{final-int}, 
with $\tilde{P}_f$ replaced
by $\tilde{P}_f^{(1\to2)^2}$ (symmetrized over parton permutations --- see Eq.~\eqref{Pftilde}).
 As in the case of the $1\to 3$ splitting-function approximation
to the planar-flow jet function, we may allow the argument scale of
$\alpha_s$ to depend on the partonic configuration. 
We show the 
ratio of the jet functions in the strongly ordered approximation to those in the basic
approximation of \sect{numeval} in Fig.~\ref{fig:1to2},  
where the scales for all factors of the strong coupling are fixed to the jet mass $m$.
The strongly ordered approximation giving rise to iterated
$1\to2$ splitting functions implies a large hierarchy between the two
splittings.  This means it is valid in the limit of small
$\Pf$ (keeping the mass fixed).
We would thus expect the two jet functions to coincide in the 
small-$\Pf$ limit.  In Fig.~\ref{fig:1to2}, the ratio does get closer
to unity at small planar flow, but a gap appears to remain.  This
gap results from the discrete number of points chosen and the use of
a linear scale; we have
verified that it closes up when going further to very small $\Pf$.
In any event, a fixed-order calculation is not valid in this
region; we should focus on the region $0.4\lesssim \Pf\lesssim 0.95$.
In this latter region, which is relevant for physics searches related to top jets for instance, the difference
between the two approximations is significant.  In this region, the
strongly ordered approximation fails to capture much of the essential physics.
This has implications for parton-shower calculations of this quantity:
we do not expect unmatched parton-shower calculations to be accurate.
Matching to tree-level matrix elements --- so long as it is done to 
a sufficiently high multiplicity as to ensure at least three matched
partons inside a jet --- will introduce the required corrections to
the strongly ordered splitting functions.  However, even the matched
calculations may be quite sensitive to the matching procedure,
and in particular to the size of the remaining region where the
pure shower calculation is used.

\subsection{Behavior at Large Planar Flow}
\label{drop}

Coming back to the $1\to 3$ computation \eqref{final-int}, let us consider the jet function at $\Pf=1$. At this point
\begin{align}
  \Delta = - \, m^4 p^2 \theta_1^2 \theta_2^2 \sin^2\!\phi \, (m^2 + 2 p^2
  \theta_1 \theta_2 \cos\phi)^2 \le 0 \ec
\label{dPf1}
\end{align}
and we are within the integration domain only when $\Delta=0$: the phase space dimension is reduced by 1. 
 Because the splitting function has no singularities within this domain,\footnote{All splitting-function singularities come from soft or collinear limits, which imply $\Pf \to 0$.} it implies that the jet function vanishes at $\Pf=1$ in our approximation. 

\begin{figure}[H]
  \centering
  \includegraphics[width=0.6\textwidth]{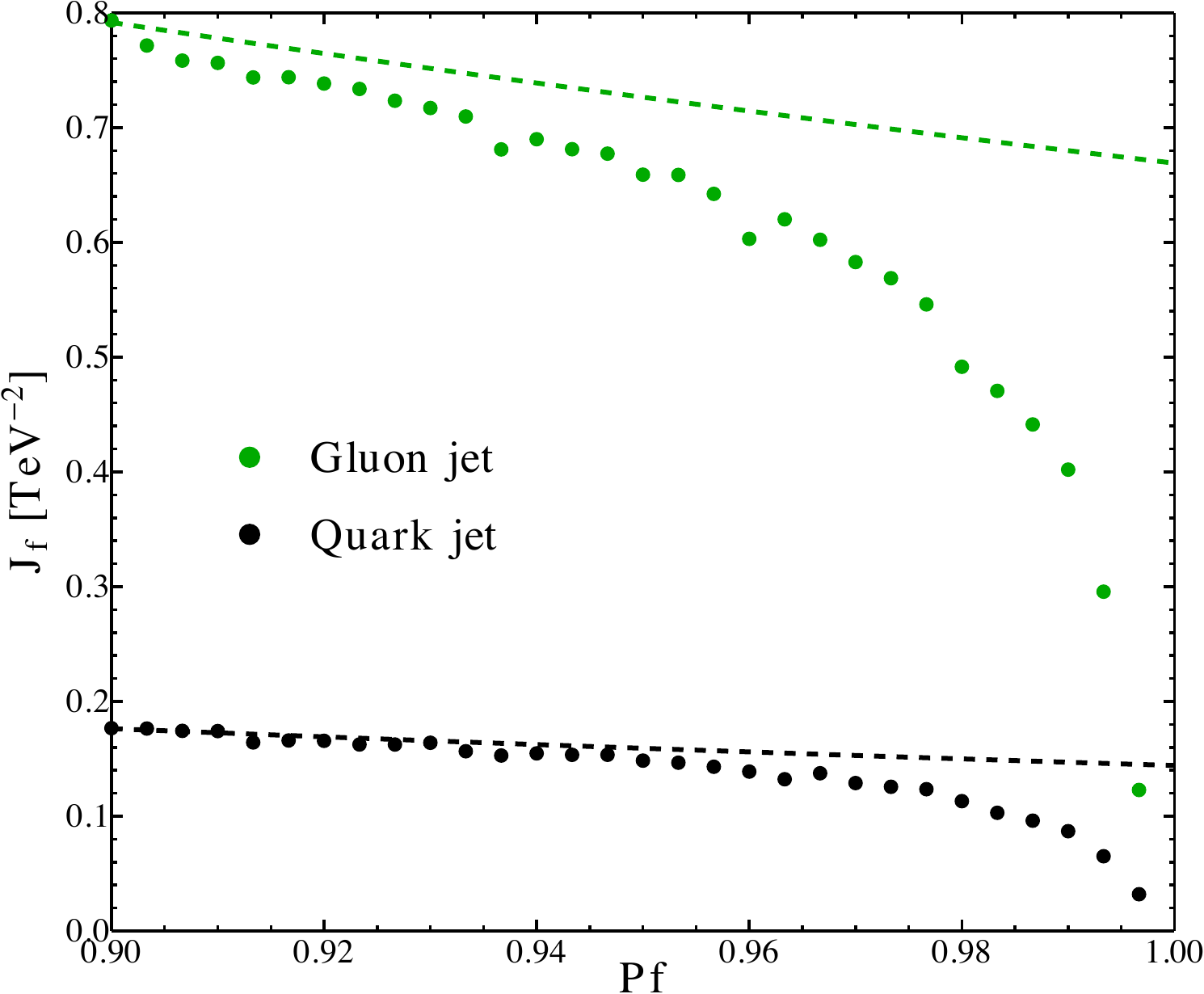}
  \caption{Semi-analytic jet function near $\Pf=1$. The dashed
  line shows the fit to the leading-log form of Eq.~\eqref{JsLL}
  with the coefficients given in Sect.~\ref{numeval}.}
  \label{fig:pfnear1}
\end{figure}

Fig.~\ref{fig:pfnear1} confirms this by showing that the semi-analytic jet function drops to zero as the planar flow approaches 1. As we will see below, the drop is a feature of our three-body approximation, and it will not be present when higher order corrections in $\alpha_s$ are included. 
It also shows the fit to the leading-log result, including the subleading coefficients in Eq.~\eqref{JsLL}, and we notice that the semi-analytic jet function diverges from the fit at the level of 10\% near $\Pf=0.95$. We heuristically take this point to mark the beginning of the drop. Our three-body approximation is not valid beyond this point.

\subsection{Comparison of Running Scales}
\label{subsec:runScale}
\begin{figure}[H]
  \centering
  \includegraphics[width=0.5\textwidth]{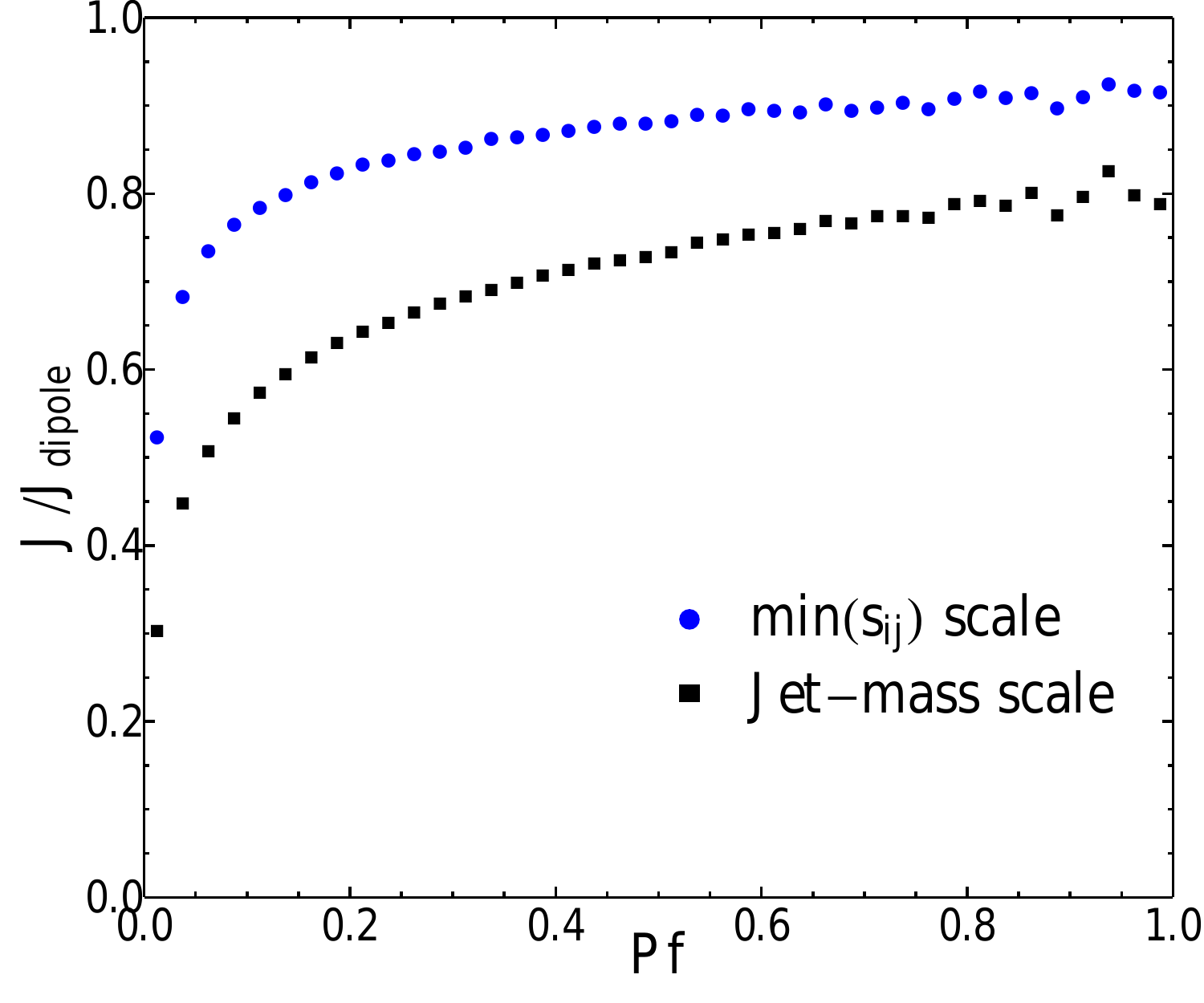}
  \caption{Ratios of gluon jet functions with different choices for $\mu$. The
  jet functions are evaluated at $p=1~\TeV$, $m=180~\GeV$ and $R=0.4$, and
  they are divided by the jet function that uses the hybrid (dipole) scale.
  }
  \label{fig:running-scales}
\end{figure}

In this section we consider how different choices of the running scale
$\mu$ affect the jet function. Recall that the jet function~\eqref{Jf1} is
of $\cO(\alpha_s^2)$, and that we evaluate the two powers of $\alpha_s$
at different scales.  One power we evaluate at the jet-mass scale (corresponding to the first $1\to2$ splitting in case of hierarchical emissions); but 
there are several possible choices of scale for the second power,
corresponding to the second (softer) $1\to2$ splitting.
 We consider three possibilities:
\begin{enumerate}
  \item Set $\mu$ to be the jet mass $m$,
  \item Set $\mu = \min_{i,j} \{ s_{ij} \}$, and
  \item Set $\mu$ to be a hybrid scale, described by the $2 \to 3$ dipole scale for gluon emission or $ s_{ij}$ in the $g\to q\bar q$ case; see
    appendix~\ref{renormalization} for details.
\end{enumerate}
We expect the last of these to be the most accurate one, and we compare
the others against it in Fig.~\ref{fig:running-scales}. The choice
of scale makes a significant difference to the value of planar-flow
jet function, of order 10--30\% at relevant values of the planar flow.
Furthermore, the ratios are not constant as a function of the planar flow.
A variation is to be expected in a leading-order calculation, as
nothing in the matrix element compensates for the change of scale.
We would expect this variation to be substantially smaller in
a next-to-leading order calculation of the jet function.  The scale
variation is hidden in parton-shower calculations (as each algorithm
chooses one particular scale), but this should be considered as an intrinsic source of
uncertainty.  Unlike the error made by applying a strongly ordered
approximation, this uncertainty is not removed by matching to 
tree-level matrix elements.  Matching to one-loop matrix elements
as well would be required to reduce it.

The results displayed in Fig.~\ref{fig:running-scales} at large $\Pf$ can be understood in a simple way. At $\Pf=1$ we have a symmetric configuration of partons, where $s_{ij} = m^2/3$. Let us assume this is the dominant configuration. For a 
three-gluon configuration (which dominates the gluon jet function), the dipole scale is then $\mu_{\mathrm{dipole}}=m/3$. We expect that the jet function ratios at $\Pf=1$ will be given by the corresponding ratios of couplings,
\begin{align}
  \frac{\as(\mathrm{min}(s_{ij}))}{\as(\mu_{\mathrm{dipole}}^2)}
  \approx 0.92
  \ecq \quad
  \frac{\as(m^2)}{\as(\mu_{\mathrm{dipole}}^2)}
  \approx 0.84
  \ed
\end{align}
This agrees nicely with Fig.~\ref{fig:running-scales}, within numerical uncertainties.

\section{Comparison to Parton-Shower Codes}
\label{sec:mc}

In this section we compare our semi-analytic result to parton-shower simulations of QCD scattering at the LHC.  The simulations are of
$2\to 4$ (matched) and $2\to 2$ (unmatched) matrix-element scattering,
plus showering. We used MadGraph/MadEvent 5 \cite{Alwall:2011uj} with
Pythia 6.420 (virtual ordered) showering \cite{Sjostrand:2006za}, and \Sherpa{}~1.3.1 \cite{Gleisberg:2008ta}, and in both cases we have used the CTEQ6L set for the parton distribution function~\cite{Pumplin:2002vw}. The
jet algorithm is anti-$k_T$ \cite{Salam:2007xv} with $R=0.4$,
implemented in FastJet \cite{Cacciari:2011ma,Cacciari:2006}. The other
parameters are $\sqrt{s} = 7~\TeV$, $950 < \pT < 1050~\GeV$ and
$|\eta|<1$. We integrate over the mass window $160 < m < 200~\GeV$,
which is consistent with our motivation for this work. The simulations
include showering but not hadronization or detector simulation. As we
show below in Sect.~\ref{hadronization}, the effect of hadronization
at large planar flow is to increase the distribution by about 15\%, while leaving the shape of the distribution
unchanged.

The semi-analytic results are computed at $p = 1~\TeV$ and integrated numerically over the same mass window. There are separate jet functions for gluon and quark jets, and the total jet function is given by $J = x J_g + (1-x) J_q$, where $x$ can be thought of as the ``fraction of gluon jets'' in the sample. While this is not a well-defined quantity, we may get a rough estimate for it by considering matrix-element $2 \to 2$ scattering. Then $x$ is given by the ratio of outgoing gluons to total outgoing partons (within the $\pT$ and $\eta$ cuts), and using this method we find $x \approx 0.24$.

The parton-shower distributions are normalized such that the integral of each over the full jet-mass and planar-flow ranges is 1. In particular, this means that the area under the plots presented below is \textit{not} 1. Our jet functions are naturally normalized in the same way, so we expect the total jet function to agree with the parton-shower result within its range of validity.

\begin{figure}[H]
  \centering
  \includegraphics[width=1\textwidth]{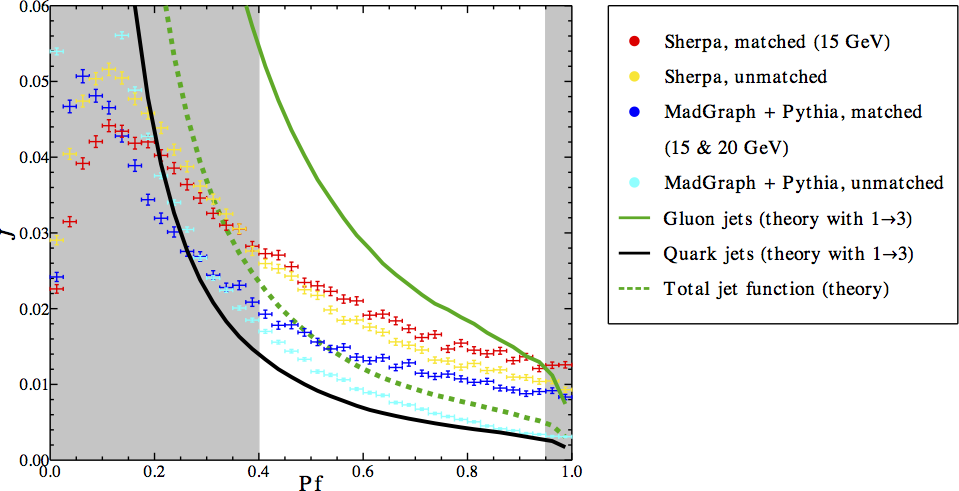}
  \caption{Jet functions from parton-shower simulations, and our predictions of  subsection~\ref{numeval} using $1\to3$ splitting functions for the quark 
and gluon planar-flow jet functions.
The upper 
(green) solid curve shows the gluon jet function, and the lower (black) 
solid curve the quark one.  The dashed (green) curve is the average jet 
function with gluon fraction $x=0.24$.
 The region of expected validity of the semi-analytic form is highlighted.
The points with error bars show the parton-shower results, in the 
highlighted region from top to bottom: matched SHERPA (red), unmatched 
SHERPA (yellow), matched MadGraph (dark blue), and unmatched MadGraph 
(light blue).  
 }
  \label{fig:pf-mc}
\end{figure}

Fig.~\ref{fig:pf-mc} shows the comparison of the semi-analytic results to the parton-shower jet functions for MadGraph (with
Pythia showering) and for \Sherpa{}. 
The second factor of the strong coupling $\alpha_s$ is evaluated at the hybrid scale as described in Sect.~\ref{subsec:runScale}.
Away from the peak we find that the parton-shower results
fall between the semi-analytic quark and gluon functions, in agreement with
the theoretical prediction.  Notice that the parton-shower jet functions display no special behavior
near $\Pf = 1$, while our semi-analytic jet functions drop to zero there (see
Sect.~\ref{drop}).  This discrepancy is due to missing higher-order contributions in the theoretical calculation. 
In detailed comparisons
with our theoretical result we will exclude this highest-$\Pf$ region,
restricting ourselves to the range $\Pf < 0.95$.

The simulations include an infrared cutoff of $15~\GeV$ at the matrix-element level, which represents the minimal $k_T$ distance between pairs of partons.\footnote{In \Sherpa{}, the CKKW matching scale $Q_{\rm cut}$ also serves as this cutoff.} A \Sherpa{} simulation with a higher cutoff of $25~\GeV$ (not shown in Fig.~\ref{fig:pf-mc}) gives a qualitatively similar result.

One can check, for example by generating random three-body jets, that a cutoff of $15~\GeV$ implies the matrix-element results apply only at $\Pf \gtrsim 0.4$. Below this only $1 \to 2$ splittings are in effect, and in addition we expect that resummation and non-perturbative effects become important. In support of this, below we will show that the effect of hadronization in the simulation does not alter the shape above $\Pf \approx 0.4$. On the theory side, as we approach the peak (at $\Pf \approx 0.1$) from above, resummation effects become important and our perturbative approximation breaks down. For the purpose of comparison we will therefore restrict ourselves to $\Pf \gtrsim 0.4\,$.

\begin{figure}[H]
  \centering
  \includegraphics[width=1\textwidth]{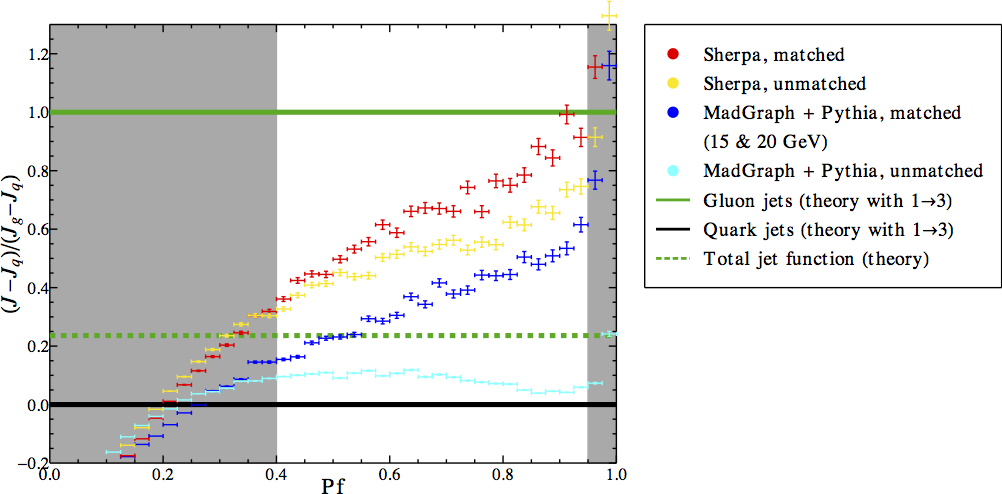}
  \caption{
  Detailed comparison between jet functions from parton-shower simulations and our semi-analytic results.
The figure shows $(J-J_q)/(J_g-J_q)$ where $J_g$ and $J_q$ are respectively the semi-analytic gluon and quark
jet functions.  These jet functions are shown by the upper (green) solid line at 1 and the lower (black) solid line at 0.  The average jet function with gluon fraction $x=0.24$ is shown by the dashed (green) line.  The region of expected validity of the semi-analytic form is highlighted.  The points with error bars show the parton-shower results, with the order and color coding as in Fig.~\ref{fig:pf-mc}\,.
}
  \label{fig:pf-mc-th}
\end{figure}

\begin{figure}[H]
  \centering
  \includegraphics[width=1\textwidth]{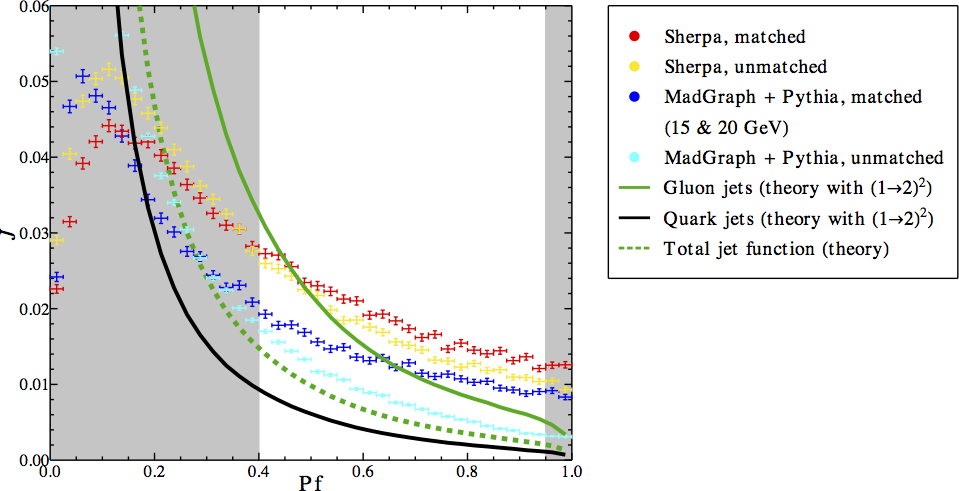}
  \caption{Comparison of jet functions from parton-shower simulations and semi-analytic results using a strong-ordering
approximation (iterated $1\rightarrow2$ splitting functions).  The two factors of the strong coupling $\alpha_s$ are evaluated at the jet-mass scale and the scale of the second splitting $s_{ij}$.  The region of expected validity of the semi-analytic form is highlighted.  The curves and points are as in Fig~ \ref{fig:pf-mc}\,.
  }
  \label{fig:pf-mc-th1to2}
\end{figure}

Fig.~\ref{fig:pf-mc-th} shows a detailed comparison of the parton-shower and theoretical jet functions. The region in which we expect to find agreement is highlighted. Finally, Fig.~\ref{fig:pf-mc-th1to2} shows the comparison to the theoretical result using the iterated $1\to 2$ splittings (see Sect.~\ref{1to2}), with the second factor of the strong coupling $\alpha_s$ evaluated at the scale of the second splitting $s_{ij}$.
 Comparing with Fig.~\ref{fig:pf-mc}, it is clear that using the $1 \to 3$ splitting function results in a significantly better approximation to the jet function.
 (The first factors of the strong coupling $\alpha_s$ are evaluated at the jet-mass scale in both cases, but the choice of second scale is different.)


\subsection{Hadronization}
\label{hadronization}

\begin{figure}[H]
  \centering
  \includegraphics[width=0.7\textwidth]{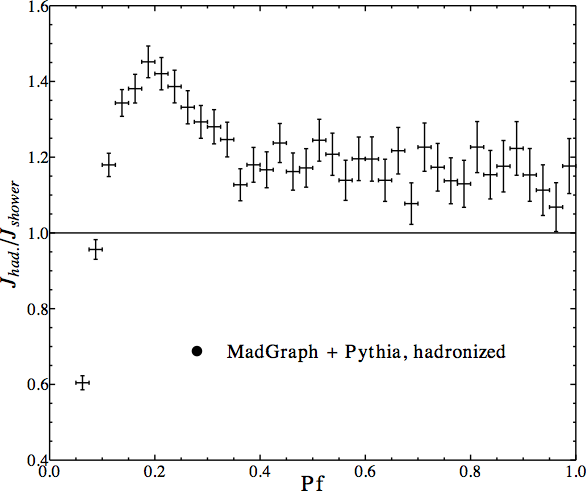}
  \caption{A comparison of parton-shower jet functions with and without
  hadronization, shown as a ratio of the two. Both
  distributions are from MadGraph + Pythia simulations with matched jets.}
  \label{fig:hadronization}
\end{figure}

The effect of turning on hadronization in the parton-shower simulation is shown in Fig.~\ref{fig:hadronization}. Below $\Pf=0.4$, where matrix-element events are discarded due to the infrared cutoff, we see that hadronization affects the shape of the jet function significantly. Above this value, hadronization affects only the overall normalization.

\subsection{Estimation of Non-Collinear Corrections}
Our theoretical computation relies on two approximations. The first is working to leading order 
in perturbation theory, which is valid
for sizeable mass and planar flow, well above the peak locations in the resummed distributions.
The validity of this approximation requires that
resummation effects be negligible, $\as(m)\log (m/pR) \ll1$ and $\as(m)\log \Pf \ll1$, for $m$ and $\Pf$ respectively.
It also requires that the values of these variables not be too close to kinematic boundaries, 
such as $m=pR$ and $\Pf=1$. 
The second approximation is the collinear approximation, valid for narrow jets.
The two approximations limit the region of validity, while offering a non-trivial window of applicability
for the calculations. The window is home to a variety of potential new-physics searches, for example those
using boosted top-quark jets.

\begin{figure}[H]
  \centering
  \subfigure[]{
    \includegraphics[width=0.45\textwidth]{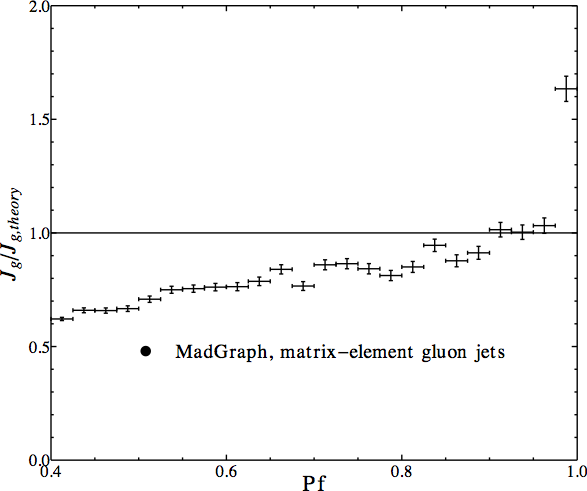}
  }
  \qquad
  \subfigure[]{
    \includegraphics[width=0.45\textwidth]{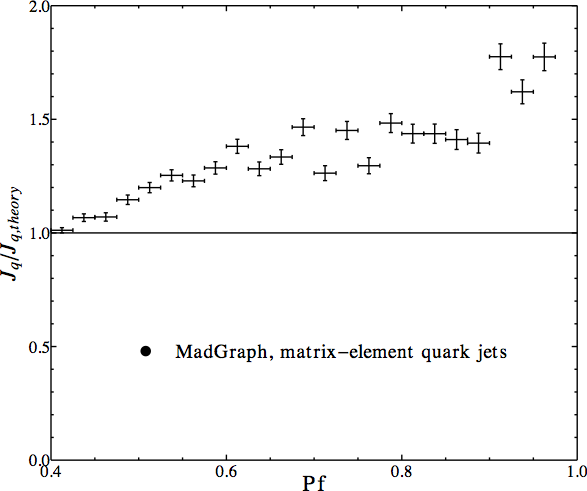}
  }
  \caption{Comparison of gluon and quark semi-analytic jet functions, with the
  results from MadGraph matrix-element $2 \to 4$ calculations. The
  parameters are $p=1~\TeV$, $m=180~\GeV$ and $R=0.4$.}
  \label{fig:me-mc-th}
\end{figure}

 We would like to have a better understanding of the corrections to our theoretical results, and accordingly we would like to separate the collinear corrections from those due to resummation.
For that purpose we consider a matrix-element calculation, with no showering or hadronization, and compare it with 
our theoretical prediction (see Fig.~\ref{fig:me-mc-th}). This calculation is carried out at leading
order in $\alpha_s$, and thus differs from our theoretical prediction only in corrections to the collinear approximation,
as the tree-level matrix elements employed are exact throughout phase space.
The difference between the two results provides an
estimation of the corrections away from the $R\rightarrow 0$ limit.
We see in Fig.~\ref{fig:me-mc-th} that these corrections vary significantly with $\Pf$. 

We normalize the exact leading-order jet function $J_f^{(e)}$
to match the way the semi-analytic jet function is normalized
at leading order $\bigl(\Ord(\as^2)\bigr)$,
so that
\begin{align}
  \int \! dm^2 J_f^{(e)} &= 
  \left(  
  \int \! d\pT d\eta \, \frac{d\sigma_{2\to f,x}}{d\pT d\eta} 
  \right)^{-1}
  \int \! d\pT d\eta dm^2 \, 
  \frac{d\sigma_{2 \to J,x}}{d\pT d\eta dm^2 d\Pf}
  \label{Jfnorm}
  \ed
\end{align}
Here, $\sigma_{2\to f,x}$ is the $2\to 2$ cross section ($x$
denotes a parton of any flavor), and $\sigma_{2 \to J,x}$ is the $2 \to
4$ cross section for an outgoing state that includes a jet $J$ of matching type. 
Both cross sections are computed to leading order in $\as$.
The integrals are over the window $950 < \pT < 1050$ GeV, $|\eta|<1$ and $160 < m < 200$ GeV, with the jet function itself evaluated at $\pT = 1$ TeV.

Let us relate these objects to quantities that are directly measurable in a Monte-Carlo integration. Consider the 
differential cross section $d\sigma$ in a given bin, for example the cross section for events  within our 
kinematic window, and with planar flow in a small range $[\Pf,\Pf+d\Pf)$. It is given by
\begin{align}
  d\sigma = \frac{dN}{N} \sigma \ec
\end{align}
where $dN$ is the number of events inside the bin, $N$ is the total number of events produced in the integration, and $\sigma$ is the total cross section computed by the simulation. Using this relation in Eq.~\eqref{Jfnorm}, we find that the jet function is given by
\begin{align}
  \int dm^2 J_f(\Pf) &= 
  \frac{1}{d\Pf} \frac{N_{\ttt}}{N_{\ttf}}
  \frac{\sigma_{\ttf}}{\sigma_{\ttt}}
  \frac{\widetilde{N}_{2 \to J, x}(\Pf)}{\tilde{N}_{2\to f,x}} \ed
\end{align}
Here, $N_{2\to X}$ and $\sigma_{2\to X}$ are the total number of events and total cross section produced in the $2\to X$ simulation; $\widetilde{N}_{2\to f,x}$ is the total number of events in the $2\to 2$ integration, with a parton of flavor $f$, that fall within our kinematic window; and $\widetilde{N}_{2 \to J, x}$ is the total number of events in the $2\to 4$ integration, with a jet of flavor $f$, that fall within our kinematic  window (including jet mass), and within our $\Pf$ bin.

\section{Conclusions}\label{sec:conc}
The planar-flow ($\Pf$) distribution of highly boosted narrow massive jets is interesting, because
 a non-vanishing value of this variable implies that the corresponding jet consists of at least three hard partons
in a perturbative description.
QCD jets with sizeable planar flow and jet mass ($m$) form an important background to various new-physics signals.
For instance, massive jets with large $\Pf$ arise in models with heavy resonances decaying dominantly to top quarks or in supersymmetric models with $R$-parity-violating gluinos.  In this paper, we have studied the planar flow
distribution of narrow QCD jets.  We obtained a semi-analytic form for this distribution, independent
of the underlying hard process giving rise to the jet.  We have made use of QCD factorization properties
to do so, and have computed jet functions which express the probability of a parent parton
fragmenting into a jet of given planar flow and mass.  We computed the leading-order approximation to these jet
functions using the universal $1\rightarrow3$ tree-level collinear splitting functions.  We compared this
approximation to a strongly ordered collinear approximation, using iterated $1\rightarrow2$ splitting
functions, and find substantial differences.  Our results are, unsurprisingly, sensitive to the choice of
scales in the strong coupling.
We have also derived
the leading-log behavior of the jet functions analytically.   Our results are expected to be valid only 
in the range $0.4\lesssim\Pf\lesssim 0.95$, and for sizeable jet mass, as fixed-order predictions will diverge
both as $\Pf\rightarrow0$ and $m\rightarrow0$.  The divergence of the planar-flow distribution should be
regulated by resummation of leading logarithms to all orders in perturbation theory.  To the best of our
knowledge, this resummation has not been computed, and the resulting resummed distribution would be
of interest.

We have compared our semi-analytic jet
function to parton-shower predictions using various event generators.  The broad
features are in agreement in the region of validity of our fixed-order calculation,
as the parton-shower results interpolate between the predicted quark and gluon jet functions.  The details
differ, however, between the parton-shower results and 
a suitably-weighted average of quark and gluon jet functions.  We find that the results from
\Sherpa{} are above our semi-analytic calculations, both with and without matching (using CKKW).  In contrast,
the results from MadGraph with Pythia showering are above our results with matching (using MLM),
but below our predictions without matching.  In the region of validity, hadronization effects increase the 
value of the jet function modestly without altering its shape.
We note that
the results of these two parton-shower codes do not agree with each other, either with or without
matching. The differences between them, even with matching, are comparable to the differences from our
semi-analytic results, and the differences are greater without matching.  This variation suggests that some caution should be
exercised when comparing these results with experimental data, and that a data-driven approach
to jet substructure should be explored as well.
The qualitative agreement of the semi-analytic results with the parton-shower calculations does however
suggest that
futher refinement of the fixed-order prediction, for example by carrying out a next-to-leading order
calculation, would be valuable.  The required $2\to4$ one-loop matrix elements are available, and have already been
used for phenomenological studies~\cite{BlackHatFourJets,BadgerFourJets}.

Planar flow is one of several three-prong substructure variables that can play a role in discriminating
highly-boosted jets arising from three-body decays of heavy particles from QCD backgrounds.  It tends to
be sensitive to soft radiation near the edge of the jet, which also makes it sensitive to pile-up.   This
motivates the use of jet filtering or template overlaps, where the hard substructure in a jet can be enhanced
in a controlled manner.  Such enhancement would be expected to bring jet functions closer to the
fixed-order perturbative ones calculated in this paper.

\section*{Acknowledgments}
It is a pleasure to thank Johan Alwall, Frank Krauss and Gavin Salam for useful discussions.
The figures for this article have been created using the LevelScheme scientific
figure preparation system \cite{levelscheme}. MF would like to thank the Weizmann Institute of Science for support while most of this work was carried out.
DAK's research is supported by the European Research Council under
Advanced Investigator Grant ERC--AdG--228301.  
GP holds the Shlomo
and Michla Tomarin development chair, supported by the
grants from Gruber foundation, IRG, ISF and Minerva.


\appendix


\section{1 $\to$ 2 Splitting Functions}
\label{sp12-func}

In this section, for completeness, we quote the $1 \to 2$ spin-averaged and color-averaged splitting 
functions~\cite{Altarelli:1977zs,Catani:1998nv} $P_{f_1f_2}$ at leading order in $\as$. Let us define
\begin{align}
  C_F = \frac{N_c^2 - 1}{2N_c} \ecq
  C_A = N_c \ecq
  T_R = \frac{1}{2} \ed
  \label{consts}
\end{align}
The functions are
\begin{align}
  P_{qg}(z) &= C_F \frac{1 + (z-1)^2}{z} \ec \\
  P_{q\bar{q}}(z) &= T_R \left[ 1 - 2z(1-z) \right] \ec \\
  P_{gg}(z) &= 2 C_A \left[ \frac{z}{1-z} + \frac{1-z}{z} + z(1-z) \right] \ec
\end{align}
and the rest are determined by charge conjugation; for example, $P_{qg} = P_{\bar{q}g}$.

\section{1 $\to$ 3  Splitting Functions}
\label{sp-func}

In this section we write down the $1\to 3$ spin-averaged and color-averaged splitting functions $P_{f_1f_2f_3}$ at leading order in $\as$. We follow the conventions of Ref.~\cite{Catani:1998nv}. Our definition of $z_i$ agrees with 
this reference in the collinear limit. In addition to Eq.~\eqref{consts}, we define
\begin{align}
  t_{ij,k} \equiv 2 \,\frac{z_i s_{jk}-z_j s_{ik}}{z_i+z_j} +
  \frac{z_i-z_j}{z_i+z_j} \,s_{ij} \ed
\end{align}

\underline{\textbf{Quark Splitting to Quarks}}
\begin{align}
  P_{{\bar Q}_1 Q_2 q_3} &= \frac{1}{2} \,
  C_F T_R \,\frac{s_{123}}{s_{12}} \left[ - \frac{t_{12,3}^2}{s_{12}s_{123}}
  +\frac{4z_3+(z_1-z_2)^2}{z_1+z_2}
  + z_1+z_2-\frac{s_{12}}{s_{123}}
  \right] \ec
  \\
  P_{{\bar q}_1q_2q_3} &=
  \left[ P_{ {\bar Q}_1 Q_2 q_3} \,
  + \,(2\leftrightarrow 3) \,\right]
  + P^{({\rm id})}_{ {\bar q}_1q_2q_3} \ec
\end{align}
where
\begin{align}
  P^{({\rm id})}_{ {\bar q}_1q_2q_3}
  &= C_F \left( C_F-\frac{1}{2} C_A \right)
  \Biggl\{ \frac{2s_{23}}{s_{12}}
  + \frac{s_{123}}{s_{12}}\Biggl(\frac{1+z_1^2}{1-z_2}-\frac{2z_2}{1-z_3}
  \Biggr)
  \notag \\ &\quad
  - \frac{s_{123}^2}{s_{12}s_{13}}\frac{z_1}{2}\left[\frac{1+z_1^2}{(1-z_2)(1-z_3)}\right] \Biggr\} + (2\leftrightarrow 3) \ed
\end{align}

\underline{\textbf{Quark Splitting to Quark + Gluons}}
\begin{align}
P_{g_1 g_2 q_3} =
C_F^2 \, P_{g_1 g_2 q_3}^{({\rm ab})}
+ \, C_F C_A P_{ g_1 g_2 q_3}^{({\rm nab})} \ec
\end{align}
where
\begin{align}
P_{g_1 g_2 q_3}^{({\rm ab})}
&=\Biggl\{\frac{s_{123}^2}{2s_{13}s_{23}}
\frac{z_3(1+z_3^2)}{z_1z_2}
+\frac{s_{123}}{s_{13}}\Biggl[\frac{z_3(1-z_1)+(1-z_2)^3}{z_1z_2}
\Biggr] -\frac{s_{23}}{s_{13}}
\Biggr\}+(1\leftrightarrow 2) \ec
\end{align}
\begin{align}
P_{g_1 g_2 q_3}^{({\rm nab})}
= &\Biggl\{\frac{t_{12,3}^2}{4s_{12}^2}+\frac{1}{4}
+\frac{s_{123}^2}{2s_{12}s_{13}}
\Biggl[\frac{(1-z_3)^2+2z_3}{z_2}
+\frac{z_2^2+2(1-z_2)}{1-z_3}\Biggr]
\notag \\ &\quad
-\frac{s_{123}^2}{4s_{13}s_{23}}z_3\Biggl[\frac{(1-z_3)^2+2z_3}{z_1z_2}
\Biggr]
+\frac{s_{123}}{2s_{12}}\Biggl[
\frac{z_1(2-2z_1+z_1^2) - z_2(6 -6 z_2+ z_2^2)}{z_2(1-z_3)}
\Biggr]
\notag \\ &\quad
+\frac{s_{123}}{2s_{13}}\Biggl[\frac{(1-z_2)^3
+z_3^2-z_2}{z_2(1-z_3)}
-\frac{z_3(1-z_1)+(1-z_2)^3}{z_1z_2}
\Biggr]\Biggr\}
+(1\leftrightarrow 2) \ed
\end{align}

\underline{\textbf{Gluon Splitting to Gluon + Quarks}}
\begin{align}
  P_{g_1q_2{\bar q}_3} =
  C_F T_R P^{({\rm ab})}_{ g_1q_2{\bar q}_3}
  + C_A T_R P^{({\rm nab})}_{ g_1q_2{\bar q}_3} \ec
\end{align}
where
\begin{align}
P^{({\rm ab})}_{ g_1q_2{\bar q}_3} &=
-2-s_{23}\left(\frac{1}{s_{12}}+\frac{1}{s_{13}}\right)
+ 2\frac{s_{123}^2}{s_{12}s_{13}}\left[1-z_1(1-z_1)-2z_2 z_3\right]
\notag \\ &\quad
-\frac{s_{123}}{s_{12}}\left(1-2z_2\right)
- \frac{s_{123}}{s_{13}}\left(1-2z_3\right)
\ec
\end{align}
\begin{align}
P^{({\rm nab})}_{ g_1q_2{\bar q}_3}
= \, &\Biggl\{-\frac{t^2_{23,1}}{4s_{23}^2}
+\frac{s_{123}^2}{2s_{13}s_{23}} z_3
\Biggl[\frac{(1-z_1)^3-z_1^3}{z_1(1-z_1)}
-\frac{2z_3\left(1-z_3 -2z_1z_2\right)}{z_1(1-z_1)}\Biggr]
\notag \\ &\quad
+\frac{s_{123}}{2s_{13}}(1-z_2)\Biggl[1
+\frac{1}{z_1(1-z_1)}-\frac{2z_2(1-z_2)}{z_1(1-z_1)}\Biggr]
\notag \\ &\quad
+\frac{s_{123}}{2s_{23}}\Biggl[\frac{1+z_1^3}{z_1(1-z_1)}
+\frac{z_1(z_3-z_2)^2-2z_2z_3(1+z_1)}
{z_1(1-z_1)}\Biggr]
\notag \\ &\quad
-\frac{1}{4}
-\frac{s_{123}^2}{2s_{12}s_{13}}\left[1-z_1(1-z_1)-2z_2z_3 \right] \Biggr\}
+ (2\leftrightarrow  3) \ed
\end{align}

\underline{\textbf{Gluon Splitting to Gluons}}
\begin{align}
P_{ g_1g_2g_3} = C_A^2 &\Biggl\{\frac{t_{12,3}^2}{4s_{12}^2}
+\frac{3}{4}+\frac{s_{123}}{s_{12}}\Biggl[4\frac{z_1z_2-1}{1-z_3}
+\frac{z_1z_2-2}{z_3}+\frac{3}{2} +\frac{5}{2}z_3
+\frac{\left(1-z_3(1-z_3)\right)^2}{z_3z_1(1-z_1)}\Biggr]
\notag \\ &\quad
+\frac{s_{123}^2}{s_{12}s_{13}}\Biggl[\frac{z_1z_2(1-z_2)(1-2z_3)}{z_3(1-z_3)}
+z_2z_3 -2 +\frac{z_1(1+2z_1)}{2}
\notag \\ &\quad
+\frac{1+2z_1(1+z_1)}{2(1-z_2)(1-z_3)}
+\frac{1-2z_1(1-z_1)}{2z_2z_3}\Biggr]\Biggr\}
+ (5\mbox{ permutations}) \ed
\end{align}

\section{Three-Parton Kinematics in the Narrow-Jet Approximation}
\label{kinematics}

In this appendix we write down kinematic quantities of three-parton
configurations, at leading order in the collinear approximation $\theta_i \le R \ll
1$, in terms of the integration variables of Eq.~\eqref{Jf}. As dictated by the kinematics we assume $m < pR$ (see Eq.~\eqref{th3-cone}), and this implies that $m \ll p$.

For parton three-momenta we use spherical coordinates $(p_i,\theta_i,\phi_i)$ with $i=1,2,3$,
relative to the jet axis $\vec{p}$, and we define $\phi\equiv\phi_1-\phi_2$.
In the collinear approximation,
\begin{align}
  E &= p + \frac{m^2}{2p} + o(R^4) \ec \\
  z_i &= \frac{p_i}{p}
  \left( 1 - \frac{m^2}{2p^2} \right) + o(R^4) \ec \\
  s_{ij} &= p_i p_j \theta_{ij}^2 + o(R^4)
  = z_i z_j p^2 \theta_{ij}^2 + o(R^4) \ec
\end{align}
where $\theta_{ij}$ is the angle between partons $i$ and $j$.

For parton 3, we have
\begin{align}
  p_3 &=
  \sqrt{(p-p_1-p_2)^2 +
  p(p_1 \theta_1^2 + p_2 \theta_2^2) - p_1 p_2 \theta_{12}^2}
  + o(R^4)
  \label{p31}
  \\ &=
  |p - p_1 - p_2|
  +
  \frac
  {p(p_1 \theta_1^2 + p_2 \theta_2^2) - p_1 p_2 \theta_{12}^2}
  {2|p - p_1 - p_2|}
  + o(R^4)
  \label{p32}
  \ec \\
  \theta_3^2 &=
  \frac{1}{|p - p_1 - p_2|} \left[
  \frac{m^2}{p}
  - p_1 \theta_1^2
  - p_2 \theta_2^2
  \right] + o(R^4)
  \label{theta3}
  \ed
\end{align}
In expanding the square root in $p_3$ we assumed that $p-p_1-p_2$ is large relative to the second term. This assumption in not valid in the limit of soft $p_3$, where in fact $p-p_1-p_2$ becomes arbitrarily small (and even negative). To avoid this limit we restrict ourselves to the range $p_1 < p_2,p_3$.

The angles $\theta_{ij}$ are given by
\begin{align}
  \theta_{12}^2 &=
  \theta_1^2 + \theta_2^2
  - 2 \theta_1 \theta_2 \cos\phi
  + o(R^4) \label{theta12} \ec\\
  \theta_{13}^2 &=
  \frac{1}{p - p_1 - p_2} \left[
  \frac{m^2}{p}
  + p \theta_1^2
  - p_2 \theta_{12}^2
  \right] + o(R^4) \label{theta13} \ec\\
  \theta_{23}^2 &=
  \frac{1}{p - p_1 - p_2} \left[
  \frac{m^2}{p}
  + p \theta_2^2
  - p_1 \theta_{12}^2
  \right] + o(R^4) \ed
  \label{theta23}
\end{align}
Finally, for the jet mass and planar flow we have
\begin{align}
  m^2 &= p \sum_{i=1}^3 p_i \theta_i^2 + o(R^4)
  = \frac{p}{p - p_1 - p_2}
  \left[
    p \left(
      p_1 \theta_1^2 + p_2 \theta_2^2
    \right)
    - p_1 p_2 \theta_{12}^2
  \right] + o(R^4)
  \ec
  \label{mcol} \\
  \Pf &= \frac{4 p^3 p_1 p_2 \theta_1^2 \theta_2^2 \sin^2\!\phi}
  {m^4 (p - p_1 - p_2)}
  + o(R^2) \ed
  \label{pfcol}
\end{align}

\section{Analytic Leading-Log Coefficients}
\label{AfAppendix}

In this section we write down the coefficients $A_f$ of the leading-log jet function \eqref{JsLL}, computed analytically from the integral \eqref{A}. We find
\begin{align}
  A_g &=
  \frac{2 \as^2 C_A^2}{\pi^2} \frac{1}{m^2}
  \log \left( \frac{p^2 R^2}{m^2} \right)
  \notag \\ &\quad
  + \frac{\as^2}{6\pi^2} 
  \frac{
  (11 C_A^2 - 4 T_R N_f C_F) (m^6 - p^6 R^6)
  + 9 C_A^2 m^2 p^2 R^2 (m^2 - p^2 R^2)
  }{m^2 (m^2 + p^2 R^2)^3}
  \ec
  \\
  A_q &=
  \frac{\as^2 C_F (C_A + C_F)}{\pi^2} \frac{1}{m^2}
  \log \left( \frac{p^2 R^2}{m^2} \right)
  + \frac{3 \as^2 C_F (C_A + C_F)}{4 \pi^2} 
  \frac{m^2 - p^2 R^2}{m^2 (m^2 + p^2 R^2)}
  \ed
\end{align}
The constants $C_A$, $C_F$ and $T_R$ are defined in Eq.~\eqref{consts}.

\section{Strong Coupling Renormalization Scales}
\label{renormalization}

In this section we explain in detail our choice of running scale $\mu$ for the second factor of $\as$ that appears in the jet function \eqref{Jf}. The first factor, corresponding to the first $1 \to 2$ splitting, is evaluated at the jet-mass scale. To make a realistic choice for $\mu$ (which corresponds to the scale of the second $1 \to 2$ splitting) we use the dipole 
model~\cite{Gustafson:1986db,Gustafson:1987rq,VinciaI,VinciaII,KraussWinter}, in which a $2 \to 3$ splitting of partons is described as an emission from a color dipole consisting of the two parent partons. The natural scale for this process is given by\footnote{In the limit of soft-collinear emission, $\mu$ becomes the $\pT$ of the emitted parton relative to its parent.}
\begin{align}
  \mu^2 = \frac{s_{12} s_{23}}{s_{123}} 
\end{align}
in the case where partons 1 and 3 form the dipole, and parton 2 is being emitted. In our case we do not know which of the partons is emitted in the second splitting, and in fact the second splitting is not even well-defined in general. We choose the scale to be the minimal one among the several options (corresponding to permutations of the partons), relying on the splitting function's preference for soft-collinear emissions.
For example, in the case of $g \to g g g$ splitting, the scale is given by
\begin{align}
  \mu_{g \to ggg}^2 =
  \min_{i,j,k} \left\{ \frac{s_{ij} s_{jk}}{s_{123}} \right\} \ed
\end{align}
The dipole model does not describe cases where the second splitting process is quark-pair production, and in such cases we choose $\mu$ to be the mass of the produced pair. The scales for the remaining processes are given by
\begin{align}
  \mu_{g \to g_1 q_2 \bar{q}_3}^2 &=
  \min \left\{
  \frac{s_{12} s_{13}}{s_{123}} \ec
  s_{23}
  \right\}
  \ec \\
  \mu_{q \to g_1 g_2 q_3}^2 &=
  \min \left\{
  \frac{s_{12} s_{23}}{s_{123}} \ec
  \frac{s_{12} s_{13}}{s_{123}}
  \right\}
  \ec \\
  \mu_{q \to \bar{q}_1 q_2 q_3}^2 &=
  \min \left\{ s_{12} \ec s_{13} \right\}
  \ec \\
  \mu_{q \to \bar{Q}_1 Q_2 q_3}^2 &= s_{12}
  \ed
\end{align}

\section{Mass and Planar-Flow Leading-Order Jet Functions from Splitting Functions}\label{sec:splitmaspf}
In this appendix we consider the jet function's behavior in the limit
of small planar flow, basically using an iterated $1\to 2$ splitting analysis in the limit of strong ordering.
Let us first examine in more detail how one can obtain the jet-mass distribution in the soft-collinear limit:
the single emission rate is proportional to,
\begin{equation}\label{split}
d\sigma_{1\to2} (z,\theta)\propto \frac{dz}{z} \frac{d\theta}{\theta}
\end{equation}
where $z$ is the energy fraction of the emitted gluon and $\theta$ is the angle between the emitted gluon and the
parent parton.
In this approximation the jet mass is 
\begin{equation}
m_J^2= E^2 z\, \theta^2
\end{equation}
For a fixed emission angle, $\theta$, we can rewrite Eq. (\ref{split}) as 
\begin{equation}
\label{splitmth}
d\sigma(m_J^2,\theta)\, \propto \frac{d m_J^2}{m_J^2} \frac{d\theta}{\theta}
\end{equation}
where now $R^2>\theta^2>  m_J^2/E^2$. We see that for a fixed mass, the distribution of $\theta$ is characterized by $1/\theta$.
The jet-mass distribution is obtained upon integrating $\theta$ between the boundaries of integration, giving
\begin{equation}
\label{splitm}
\frac{d\sigma_{1\to2}}{dm_J^2}\, \propto \frac{1}{m_J^2} \log\left(\frac{R^2 E^2}{m_J^2}\right) 
\end{equation}
where in hadronic collisions $E$ should be replaced by $\pT$.  The missing proportionality coefficient is nothing but $\alpha_s \,C_A /\pi$ (for a gluon jet).

Next let us try to obtain an expression for the planar-flow distribution in the limit of small planar flow for a massive jet. 
In this limit, we expect that the dominant contribution arises from configurations where the third parton is soft and collinear to either of the first two partons.
This contribution should be described well by a $1\to2$ splitting function.
We can thus iterate the above expression starting with a two parton configuration of mass $m_J$ where the two patrons are separated by an angle $\sim \theta$.
We can now add a second emission with an angle $\theta'$ such that $\theta' \ll \theta$ and an energy fraction $z' \ll z$ such that the third parton can be thought as soft and collinear with respect to either of the first two partons. The 
differential cross section describing this configuration is given by,
\begin{equation}
\label{splitmth2}
d\sigma(m_J^2,\theta)\times d\sigma_{1\to2}(\theta', z')\, \propto \frac{d m_J^2}{m_J^2} 
\frac{d\theta}{\theta}\times \frac{dz'}{z'} \frac{d\theta'}{\theta'}
\end{equation}
where now $z'$ is integrated between 0 and $z/2$ at most. (At the upper boundary, both partons have the same energy; this violates the soft approximation but is highly suppressed due to the weight in the splitting function.) 
Likewise, $\theta'$ is integrated between 0 and $\theta$ at most.
For a fixed value of mass and planar flow $z'$ and $\theta'$ are not really independent to leading order. 
The tensor $\cI$ describing this configuration is
\begin{align}
  \cI \; &=
  \! E
  \begin{pmatrix}
    [z,\,(1-z)] z'   \theta'^2 \sin^2 \phi  & [z\theta_s,\,(1-z)(1-\theta_s)] z'   \theta' \sin 2\phi /2   \\
     [z\theta_s,\,(1-z)(1-\theta_s)] z'   \theta' \sin 2\phi /2  & (1-z)(\theta- \theta_s)^2+z\theta_s^2
  \end{pmatrix}
  \ed
\end{align}
It is easy to check that $ \theta_s=(1-z) \theta$ and $(1-z)(\theta- \theta_s)^2+z\theta_s^2=z(1-z)\theta^2=m^2/E^2$, and the bracketed expressions in the (11,12,21) entries cover the cases where the third parton is emitted from the harder or softer of the first two patrons, the softer one being characterized by an angle $\theta_s=(1-z) \theta$. $\phi$ describes the azimuthal angle --- the third parton's emission angle relative to the line connecting the two hard partons.
To leading order in $z'$ we find
\begin{equation}
{\rm Pf}= 4\frac{E^2}{m^2} \,[z,\,(1-z)] z'   \theta'^2 \sin^2 \phi \approx  4\frac{E^2}{m^2} \,z'   \theta'^2 \sin^2 \phi  \ec
\end{equation}
where on the right-hand side of the above relation we have focused on the most singular region $z'\ll z\ll1$.
Similarly to the jet-mass case we can interchange  $z'$ and Pf in the singular region, Pf$\,\ll1$.
Integrating over the first emission and using the splitting function for both radiations (assuming only gluons for simplicity), focusing on the most singular region $z'\ll z\ll1,$ $\theta'\ll\theta\ll1$ and changing variables from $z'$ to $\Pf$ (we can ignore $\phi$ as in this approximation the splitting function does not depend on it) we find
\begin{equation}
\left. \frac{d^3\sigma}{d m d {\rm Pf} d\theta'} \right|_{\rm Pf\ll1}
\approx   \frac{ \alpha_s \,C_A}{\pi m_J^2} 
\log\left(\frac{R^2 E^2}{m_J^2}\right) \times \frac{ \alpha_s \,C_A}{\pi} \frac{1}{\rm Pf} \frac{1}{\theta'} \ed
\end{equation}
For a given Pf the range for $\theta'$ is 
\begin{equation}
\sqrt{\rm Pf}\,
\frac{ 2 m}{E } \ll\theta'\ll R\,.
\end{equation}
Integrating over $\theta'$ yields the final expression for the Pf distribution for jets of small Pf and mass,
\begin{equation}
\left(\frac{d^2\sigma}{d {\rm Pf} d m}\right)_{\rm Pf\ll1}\approx 
\frac{ \alpha_s^2 \,C_A^2}{\pi^2 m_J^2} \log\left(\frac{R^2 E^2}{m_J^2}\right)\times
\frac{1}{ {\rm Pf}}\,\log\left( \frac{E R}{ { 2 m \sqrt{\rm Pf}}}\right)
\end{equation}
where in hadronic collisions $E$ should be replaced with $\pT$.


\begin{thebibliography}{99}

\bibitem{:2012txa}
  G.~Aad {\it et al.}  [ATLAS Collaboration],
  ``A search for $t\bar{t}$ resonances in lepton+jets events with highly boosted top quarks collected in $pp$ collisions at $\sqrt{s} = 7$ TeV with the ATLAS detector,''
  JHEP {\bf 1209}, 041 (2012)
  [arXiv:1207.2409 [hep-ex]]; \\
   G.~Aad {\it et al.}  [ATLAS Collaboration],
  ``Search for pair production of heavy top-like quarks decaying to a high-pT $W$ boson and a $b$ quark in the lepton plus jets final state at $\sqrt{s}=7$ TeV with the ATLAS detector,''
  arXiv:1210.5468 [hep-ex]; \\
  G.~Aad {\it et al.}  [ATLAS Collaboration],
  ``Search for resonances decaying into top-quark pairs using fully hadronic decays in pp collisions with ATLAS at sqrt(s) = 7 TeV,''
  arXiv:1211.2202 [hep-ex].
  
\bibitem{Chatrchyan:2012cx}
  S.~Chatrchyan {\it et al.}  [CMS Collaboration],
  ``Search for resonant $t\bar{t}$ production in lepton+jets events in $pp$ collisions at $\sqrt{s}=7$ TeV,''
  [arXiv:1209.4397 [hep-ex]]; \\
   S.~Chatrchyan {\it et al.}  [CMS Collaboration],
  ``Search for anomalous t t-bar production in the highly-boosted all-hadronic final state,''
  JHEP {\bf 1209}, 029 (2012)
  [arXiv:1204.2488 [hep-ex]]; \\
  CMS collaboration, confnote, EXO-11-006.
  
\bibitem{Aad:2012meb}
  G.~Aad {\it et al.}  [ATLAS Collaboration],
  ``ATLAS measurements of the properties of jets for boosted particle searches,''
  arXiv:1206.5369 [hep-ex]; \\
 G.~Aad {\it et al.}  [ATLAS Collaboration],
  ``Jet mass and substructure of inclusive jets in $\sqrt{s}=7$ TeV $pp$ collisions with the ATLAS experiment,''
  JHEP {\bf 1205}, 128 (2012)
  [arXiv:1203.4606 [hep-ex]].

\bibitem{Rappoccio:2012zz}
  S.~Rappoccio [CMS Collaboration],
  ``Jets and jet substructure,''
  AIP Conf.\ Proc.\  {\bf 1441}, 820 (2012);
 CMS Collaboration,
  ``Jet Substructure Algorithms,''
  CMS-PAS-JME-10-013.

\bibitem{Aaltonen:2011pg}
  T.~Aaltonen {\it et al.}  [CDF Collaboration],
  ``Study of Substructure of High Transverse Momentum Jets Produced in Proton-Antiproton Collisions at $\sqrt{s}=1.96$ TeV,''
  Phys.\ Rev.\ D {\bf 85}, 091101 (2012)
  [arXiv:1106.5952 [hep-ex]]; \\
 T.~Aaltonen {\it et al.}  [CDF Collaboration], CDF/ANAL/TOP/ PUBLIC/10234.
  
\bibitem{Butterworth:2002tt}
  J.~M.~Butterworth, B.~E.~Cox, J.~R.~Forshaw,
  ``$W W$ scattering at the CERN LHC,''
  Phys.\ Rev.\  {\bf D65 } (2002)  096014
  [hep-ph/0201098].

\bibitem{Agashe:2006hk}
  K.~Agashe, A.~Belyaev, T.~Krupovnickas, G.~Perez, J.~Virzi,
  ``LHC Signals from Warped Extra Dimensions,''
  Phys.\ Rev.\  {\bf D77 } (2008)  015003
  [hep-ph/0612015].

\bibitem{Lillie:2007yh}
  B.~Lillie, L.~Randall, L.-T.~Wang,
  ``The Bulk RS KK-gluon at the LHC,''
  JHEP {\bf 0709 } (2007)  074
  [hep-ph/0701166].

\bibitem{Butterworth:2008iy}
  J.~M.~Butterworth, A.~R.~Davison, M.~Rubin, G.~P.~Salam,
  ``Jet substructure as a new Higgs search channel at the LHC,''
  Phys.\ Rev.\ Lett.\  {\bf 100 } (2008)  242001
  [arXiv:0802.2470 [hep-ph]].
   
\bibitem{Butterworth:2009qa}
  J.~M.~Butterworth, J.~R.~Ellis, A.~R.~Raklev, G.~P.~Salam,
  ``Discovering baryon-number violating neutralino decays at the LHC,''
  Phys.\ Rev.\ Lett.\  {\bf 103 } (2009)  241803
  [arXiv:0906.0728 [hep-ph]].

\bibitem{Butterworth:2007ke}
  J.~M.~Butterworth, J.~R.~Ellis, A.~R.~Raklev,
  ``Reconstructing sparticle mass spectra using hadronic decays,''
  JHEP {\bf 0705 } (2007)  033
  [hep-ph/0702150 [hep-ph]].

\bibitem{Ellis:2007ib}
S.~D.~Ellis, J.~Huston, K.~Hatakeyama, P.~Loch and M.~Tonnesmann,
  ``Jets in hadron-hadron collisions,''
  Prog.\ Part.\ Nucl.\ Phys.\  {\bf 60} (2008) 484
  [arXiv:0712.2447 [hep-ph]].

\bibitem{Abdesselam:2010pt}
  A.~Abdesselam, E.~B.~Kuutmann, U.~Bitenc, G.~Brooijmans, J.~Butterworth, P.~Bruckman de Renstrom, D.~Buarque Franzosi, R.~Buckingham {\it et al.},
  ``Boosted objects: A Probe of beyond the Standard Model physics,''
  Eur.\ Phys.\ J.\  {\bf C71 } (2011)  1661.
  [arXiv:1012.5412 [hep-ph]]; \\
   A.~Altheimer, S.~Arora, L.~Asquith, G.~Brooijmans, J.~Butterworth, M.~Campanelli, B.~Chapleau and A.~E.~Cholakian {\it et al.},
  ``Jet Substructure at the Tevatron and LHC: New results, new tools, new benchmarks,''
  J.\ Phys.\ G {\bf 39} (2012) 063001
  [arXiv:1201.0008 [hep-ph]].

\bibitem{Salam:2009jx}
  G.~P.~Salam,
  ``Towards Jetography,''
  Eur.\ Phys.\ J.\  {\bf C67 } (2010)  637-686
  [arXiv:0906.1833 [hep-ph]].

\bibitem{Nath:2010zj}
  P.~Nath, B.~D.~Nelson, H.~Davoudiasl, B.~Dutta, D.~Feldman, Z.~Liu, T.~Han, P.~Langacker {\it et al.},
  ``The Hunt for New Physics at the Large Hadron Collider,''
  Nucl.\ Phys.\ Proc.\ Suppl.\  {\bf 200-202 } (2010)  185-417
  [arXiv:1001.2693 [hep-ph]].

\bibitem{Almeida:2011ud}
  L.~G.~Almeida, R.~Alon and M.~Spannowsky,
  ``Structure of Fat Jets at the Tevatron and Beyond,''
  Eur.\ Phys.\ J.\ C {\bf 72}, 2113 (2012)
  [arXiv:1110.3684 [hep-ph]] (part of special review on ``Top and flavour physics in the LHC era'', Eds.   A.~J.~Buras, G.~Perez, T.~A.~Schwarz and T.~M.~P.~Tait,
  Eur.\ Phys.\ J.\ C {\bf 72}, 2105 (2012)); \\
  T.~Plehn and M.~Spannowsky,
  ``Top Tagging,''
  J.\ Phys.\ G {\bf 39}, 083001 (2012)
  [arXiv:1112.4441 [hep-ph]].

\bibitem{GurAri:2011vx}
  G.~Gur-Ari, M.~Papucci and G.~Perez,
  ``Classification of Energy Flow Observables in Narrow Jets,''
  arXiv:1101.2905 [hep-ph].

\bibitem{Almeida:2008yp}
  L.~G.~Almeida, S.~J.~Lee, G.~Perez, G.~F.~Sterman, I.~Sung, J.~Virzi,
  ``Substructure of High-$\pT$ Jets at the LHC,''
  Phys.\ Rev.\  {\bf D79 } (2009)  074017
  [arXiv:0807.0234 [hep-ph]].

\bibitem{Almeida:2010pa}
  L.~G.~Almeida, S.~J.~Lee, G.~Perez, G.~Sterman and I.~Sung,
  ``Template Overlap Method for Massive Jets,''
  Phys.\ Rev.\ D {\bf 82}, 054034 (2010)
  [arXiv:1006.2035 [hep-ph]].

\bibitem{Berger:2003iw} 
  C.~F.~Berger, T.~Kucs and G.~F.~Sterman,
  ``Event shape / energy flow correlations,''
  Phys.\ Rev.\ D {\bf 68}, 014012 (2003)
  [hep-ph/0303051].

\bibitem{Brooijmans:2010tn}
  G.~Brooijmans {\it et al.}  [New Physics Working Group Collaboration],
  ``New Physics at the LHC. A Les Houches Report: Physics at TeV Colliders 2009 - New Physics Working Group,''
  arXiv:1005.1229 [hep-ph]; \\
Y.~Eshel, O.~Gedalia, G.~Perez and Y.~Soreq,
  ``Implications of the Measurement of Ultra-Massive Boosted Jets at CDF,''
  Phys.\ Rev.\ D {\bf 84} (2011) 011505
  [arXiv:1101.2898 [hep-ph]].

\bibitem{Almeida:2008tp}
  L.~G.~Almeida, S.~J.~Lee, G.~Perez, I.~Sung, J.~Virzi,
  ``Top Jets at the LHC,''
  Phys.\ Rev.\  {\bf D79 } (2009)  074012
  [arXiv:0810.0934 [hep-ph]].

\bibitem{Thaler:2008ju}
  J.~Thaler, L.~-T.~Wang,
  ``Strategies to Identify Boosted Tops,''
  JHEP {\bf 0807 } (2008)  092
  [arXiv:0806.0023 [hep-ph]].
  
\bibitem{Cacciari:2008gn}
  M.~Cacciari, G.~P.~Salam and G.~Soyez,
  ``The Catchment Area of Jets,''
  JHEP {\bf 0804} (2008) 005
  [arXiv:0802.1188 [hep-ph]].

\bibitem{Krohn:2009wm}
D.~Krohn, J.~Shelton and L.~-T.~Wang,
  ``Measuring the Polarization of Boosted Hadronic Tops,''
  JHEP {\bf 1007}, 041 (2010)
  [arXiv:0909.3855 [hep-ph]].

\bibitem{Soyez:2012hv}
  G.~Soyez, G.~P.~Salam, J.~Kim, S.~Dutta and M.~Cacciari,
  ``Pileup subtraction for jet shapes,''
  arXiv:1211.2811 [hep-ph]; \\
  R.~Alon, E.~Duchovni, G.~Perez, A.~P.~Pranko and P.~K.~Sinervo,
  ``A Data-driven method of pile-up correction for the substructure of massive jets,''
  Phys.\ Rev.\ D {\bf 84}, 114025 (2011)
  [arXiv:1101.3002 [hep-ph]]; \\
 S.~Sapeta, Q.~C.~Zhang and Q.~C.~Zhang,
  ``The mass area of jets,''
  JHEP {\bf 1106}, 038 (2011)
  [arXiv:1009.1143 [hep-ph]].
  
\bibitem{Krohn:2009th}
  D.~Krohn, J.~Thaler and L.~-T.~Wang,
  ``Jet Trimming,''
  JHEP {\bf 1002}, 084 (2010)
  [arXiv:0912.1342 [hep-ph]]; \\
S.~D.~Ellis, C.~K.~Vermilion and J.~R.~Walsh,
  ``Recombination Algorithms and Jet Substructure: Pruning as a Tool for Heavy Particle Searches,''
  Phys.\ Rev.\ D {\bf 81}, 094023 (2010)
  [arXiv:0912.0033 [hep-ph]].

\bibitem{JetSubstructuresRefs1}
  J.~Thaler and K.~Van Tilburg,
  ``Maximizing Boosted Top Identification by Minimizing N-subjettiness,''
  JHEP {\bf 1202}, 093 (2012)
  [arXiv:1108.2701 [hep-ph]]; \\
  C.~Chen,
  ``New approach to identifying boosted hadronically-decaying particle using jet substructure in its center-of-mass frame,''
  Phys.\ Rev.\ D {\bf 85}, 034007 (2012)
  [arXiv:1112.2567 [hep-ph]]; \\
  Z.~Han,
  ``Tracking the Identities of Boosted Particles,''
  Phys.\ Rev.\ D {\bf 86}, 014026 (2012)
  [arXiv:1112.3378 [hep-ph]]; \\
  I.~Feige, M.~Schwartz, I.~Stewart and J.~Thaler,
  ``Precision Jet Substructure from Boosted Event Shapes,''
  Phys.\ Rev.\ Lett.\  {\bf 109}, 092001 (2012)
  [arXiv:1204.3898 [hep-ph]]; \\
  G.~Brooijmans,
  ``High $p_{\rm T}$ hadronic top quark identification. Part I: Jet mass and Ysplitter,''
  ATL-PHYS-CONF-2008-008;
  T.~Plehn, G.~P.~Salam and M.~Spannowsky,
  ``Fat Jets for a Light Higgs,''
  Phys.\ Rev.\ Lett.\  {\bf 104}, 111801 (2010)
  [arXiv:0910.5472 [hep-ph]]; \\
   T.~Plehn, M.~Spannowsky and M.~Takeuchi,
  ``How to Improve Top Tagging,''
  Phys.\ Rev.\ D {\bf 85}, 034029 (2012)
  [arXiv:1111.5034 [hep-ph]]; \\
   T.~Plehn, M.~Spannowsky and M.~Takeuchi,
  ``Boosted Semileptonic Tops in Stop Decays,''
  JHEP {\bf 1105}, 135 (2011)
  [arXiv:1102.0557 [hep-ph]].
  
\bibitem{JetSubstructuresRefs2}
  J.~Thaler and K.~Van Tilburg,
  ``Identifying Boosted Objects with N-subjettiness,''
  JHEP {\bf 1103}, 015 (2011)
  [arXiv:1011.2268 [hep-ph]]; \\
  Y.~Cui, Z.~Han and M.~D.~Schwartz,
  ``W-jet Tagging: Optimizing the Identification of Boosted Hadronically-Decaying W Bosons,''
  Phys.\ Rev.\ D {\bf 83}, 074023 (2011)
  [arXiv:1012.2077 [hep-ph]]; \\
  C.~Delaunay, O.~Gedalia, S.~J.~Lee, G.~Perez and E.~Ponton,
  ``Extraordinary Phenomenology from Warped Flavor Triviality,''
  Phys.\ Lett.\ B {\bf 703}, 486 (2011)
  [arXiv:1101.2902 [hep-ph]].
  
\bibitem{JetSubstructuresRefs3}
  A.~Hook, M.~Jankowiak and J.~G.~Wacker,
  ``Jet Dipolarity: Top Tagging with Color Flow,''
  JHEP {\bf 1204}, 007 (2012)
  [arXiv:1102.1012 [hep-ph]]; \\
  D.~E.~Soper and M.~Spannowsky,
  ``Finding physics signals with shower deconstruction,''
  Phys.\ Rev.\ D {\bf 84}, 074002 (2011)
  [arXiv:1102.3480 [hep-ph]]; arXiv:1211.3140 [hep-ph]; \\
  M.~Jankowiak and A.~J.~Larkoski,
  ``Jet Substructure Without Trees,''
  JHEP {\bf 1106}, 057 (2011)
  [arXiv:1104.1646 [hep-ph]]; \\
  S.~D.~Ellis, C.~K.~Vermilion and J.~R.~Walsh,
  ``Recombination Algorithms and Jet Substructure: Pruning as a Tool for Heavy Particle Searches,''
  Phys.\ Rev.\ D {\bf 81}, 094023 (2010)
  [arXiv:0912.0033 [hep-ph]]; \\
  S.~D.~Ellis, C.~K.~Vermilion, J.~R.~Walsh, A.~Hornig and C.~Lee,
  ``Jet Shapes and Jet Algorithms in SCET,''
  JHEP {\bf 1011}, 101 (2010)
  [arXiv:1001.0014 [hep-ph]]; \\
  A.~Banfi, M.~Dasgupta, K.~Khelifa-Kerfa and S.~Marzani,
  ``Non-global logarithms and jet algorithms in high-pT jet shapes,''
  JHEP {\bf 1008}, 064 (2010)
  [arXiv:1004.3483 [hep-ph]]; \\
  D.~E.~Kaplan, K.~Rehermann, M.~D.~Schwartz and B.~Tweedie,
  ``Top Tagging: A Method for Identifying Boosted Hadronically Decaying Top Quarks,''
  Phys.\ Rev.\ Lett.\  {\bf 101}, 142001 (2008)
  [arXiv:0806.0848 [hep-ph]].
  
\bibitem{JetSubstructuresRefs4}
  A.~Hornig, C.~Lee and G.~Ovanesyan,
  ``Effective Predictions of Event Shapes: Factorized, Resummed, and Gapped Angularity Distributions,''
  JHEP {\bf 0905}, 122 (2009)
  [arXiv:0901.3780 [hep-ph]]; \\
  D.~Krohn, J.~Thaler and L.~-T.~Wang,
  ``Jets with Variable R,''
  JHEP {\bf 0906}, 059 (2009)
  [arXiv:0903.0392 [hep-ph]]; \\
  S.~D.~Ellis, C.~K.~Vermilion and J.~R.~Walsh,
  ``Techniques for improved heavy particle searches with jet substructure,''
  Phys.\ Rev.\ D {\bf 80}, 051501 (2009)
  [arXiv:0903.5081 [hep-ph]]; \\
  C.~Lee, A.~Hornig and G.~Ovanesyan,
  ``Probing the Structure of Jets: Factorized and Resummed Angularity Distributions in SCET,''
  PoS EFT {\bf 09}, 010 (2009)
  [arXiv:0905.0168 [hep-ph]]; \\
  M.~Cacciari,
  ``Recent Progress in Jet Algorithms and Their Impact in Underlying Event Studies,''
  arXiv:0906.1598 [hep-ph].

\bibitem{Alon:2011}
  CDF Collaboration, CDF/ANAL/TOP/ PUBLIC/10234.

\bibitem{Altarelli:1977zs}
  G.~Altarelli, G.~Parisi,
  ``Asymptotic Freedom in Parton Language,''
  Nucl.\ Phys.\  {\bf B126 } (1977)  298.

\bibitem{Sterman:2004pd}
  G.~F.~Sterman,
 ``QCD and Jets,'' hep-ph/0412013.

\bibitem{Dremin:2005kn}
  I.~M.~Dremin,
  ``Soft and hard Jets in QCD,''
  AIP Conf.\ Proc.\  {\bf 828 } (2006)  30-34
  [hep-ph/0510250].

\bibitem{Han:2005mu}
  T.~Han,
  ``Collider phenomenology: Basic knowledge and techniques,''
  [hep-ph/0508097].

\bibitem{Sterman:1995fz}
  G.~F.~Sterman,
  ``Partons, factorization and resummation, TASI 95,''
  [hep-ph/9606312].

\bibitem{Catani:1998nv}
  S.~Catani, M.~Grazzini,
  ``Collinear factorization and splitting functions for next-to-next-to-leading order QCD calculations,''
  Phys.\ Lett.\  {\bf B446 } (1999)  143-152
  [hep-ph/9810389].

\bibitem{Peskin:1995ev}
  M.~E.~Peskin, D.~V.~Schroeder,
  ``An Introduction to quantum field theory,''
  Reading, USA: Addison-Wesley (1995).
  
\bibitem{Campbell:1997hg}
  J.~M.~Campbell, E.~W.~N.~Glover,
  ``Double unresolved approximations to multiparton scattering amplitudes,''
  Nucl.\ Phys.\  {\bf B527 } (1998)  264-288
  [hep-ph/9710255].
  
\bibitem{Catani:1992ua} 
  S.~Catani, L.~Trentadue, G.~Turnock and B.~R.~Webber,
  Nucl.\ Phys.\ B {\bf 407}, 3 (1993).



\bibitem{Alwall:2011uj}
  J.~Alwall, M.~Herquet, F.~Maltoni, O.~Mattelaer, T.~Stelzer,
  ``MadGraph 5 : Going Beyond,''
  JHEP {\bf 1106 } (2011)  128
  [arXiv:1106.0522 [hep-ph]].

\bibitem{Sjostrand:2006za}
  T.~Sjostrand, S.~Mrenna and P.~Z.~Skands,
  ``PYTHIA 6.4 Physics and Manual,''
  JHEP {\bf 0605}, 026 (2006)
  [hep-ph/0603175].

\bibitem{Gleisberg:2008ta}
  T.~Gleisberg, S.~.Hoeche, F.~Krauss, M.~Schonherr, S.~Schumann, F.~Siegert, J.~Winter,
  ``Event generation with SHERPA 1.1,''
  JHEP {\bf 0902 } (2009)  007
  [arXiv:0811.4622 [hep-ph]].
  
\bibitem{Pumplin:2002vw} 
  J.~Pumplin, D.~R.~Stump, J.~Huston, H.~L.~Lai, P.~M.~Nadolsky and W.~K.~Tung,
  ``New generation of parton distributions with uncertainties from global QCD analysis,''
  JHEP {\bf 0207}, 012 (2002)
  [hep-ph/0201195].


\bibitem{Salam:2007xv}
  G.~P.~Salam, G.~Soyez,
  ``A practical Seedless Infrared-Safe Cone jet algorithm,''
  JHEP {\bf 0705}, 086 (2007)
  [arXiv:0704.0292 [hep-ph]].

\bibitem{Cacciari:2011ma}
  M.~Cacciari, G.~P.~Salam and G.~Soyez,
  ``FastJet user manual,''
  Eur.\ Phys.\ J.\ C {\bf 72} (2012) 1896
  [arXiv:1111.6097 [hep-ph]].

\bibitem{Cacciari:2006}
  M.~Cacciari and G.~P.~Salam,
  ``Dispelling the $N^{3}$ myth for the $k_t$ jet-finder,''
  Phys.\ Lett.\ B\ {\bf 641} (2006) 57
  [hep-ph/0512210].

\bibitem{BlackHatFourJets}
Z.~Bern, G.~Diana, L.~J.~Dixon, F.~Febres Cordero, S.~Hoeche, D.~A.~Kosower, H.~Ita, D.~Ma\^{\i}tre, K.~Ozeren,
  ``Four-Jet Production at the Large Hadron Collider at Next-to-Leading Order in QCD,''
Phys.\ Rev.\ Lett.\  {\bf 109}, 042001 (2012)  [arXiv:1112.3940 [hep-ph]].  

\bibitem{BadgerFourJets}
S.~Badger, B.~Biedermann, P.~Uwer and V.~Yundin,
  ``NLO QCD corrections to multi-jet production at the LHC with a centre-of-mass energy of $\sqrt{s}=8$ TeV,''
arXiv:1209.0098 [hep-ph].  

\bibitem{levelscheme}
  M.~A.~Caprio,
  Comput. Phys. Commun.  171, 107 (2005),
  \url{http://scidraw.nd.edu/levelscheme}.

\bibitem{Gustafson:1986db}
  G.~Gustafson,
  ``Dual Description of a Confined Color Field,''
  Phys.\ Lett.\ B {\bf 175}, 453 (1986).

\bibitem{Gustafson:1987rq}
  G.~Gustafson and U.~Pettersson,
  ``Dipole Formulation of QCD Cascades,''
  Nucl.\ Phys.\ B {\bf 306}, 746 (1988).

\bibitem{VinciaI}
W.~T.~Giele, D.~A.~Kosower and P.~Z.~Skands,
  ``A Simple shower and matching algorithm,''  
Phys.\ Rev.\ D {\bf 78}, 014026 (2008)  [arXiv:0707.3652 [hep-ph]].
\bibitem{VinciaII}
  W.~T.~Giele, D.~A.~Kosower and P.~Z.~Skands,
  ``Higher-Order Corrections to Timelike Jets,''  
Phys.\ Rev.\ D {\bf 84}, 054003 (2011)  [arXiv:1102.2126 [hep-ph]].  

\bibitem{KraussWinter}
J.~-C.~Winter and F.~Krauss,
``Initial-state showering based on colour dipoles connected to incoming parton lines,''  
JHEP {\bf 0807}, 040 (2008)  [arXiv:0712.3913 [hep-ph]].  



\end{thebibliography}
\end{document}